\documentclass[12pt]{article}
\usepackage{amsthm,amsmath,amssymb,bm,bbm,graphicx,natbib,color}
\usepackage{enumerate}

\usepackage[ruled]{algorithm2e}
\usepackage{multirow}
\usepackage{float}
\usepackage{url} 
\usepackage[colorlinks=true, citecolor=blue]{hyperref}
\usepackage{multirow,booktabs}

\graphicspath{{figures/}}

\newcommand{\blind}{1}

\usepackage{subcaption}
\usepackage{xcolor}
\usepackage{soul}
\usepackage{booktabs}

\usepackage{mathtools}

\newcommand{\wh}{\widehat}
\newcommand{\wt}{\widetilde}
\newcommand{\rmd}{\mathrm{d}}

\DeclareMathOperator*{\argmin}{arg\,min}

\newcommand{\bc}{\mathbf{c}}
\newcommand{\bd}{\mathbf{d}}

\newcommand{\bh}{\mathbf{h}}
\newcommand{\bH}{\mathbf{H}}

\newcommand{\bbeta}{\bm{\beta}}
\newcommand{\bphi}{\bm{\phi}}
\newcommand{\bpsi}{\bm{\psi}}

\newcommand{\btheta}{\bm{\theta}}
\newcommand{\bgamma}{\bm{\gamma}}

\numberwithin{equation}{section}

\newcommand{\ra}[1]{\renewcommand{\arraystretch}{#1}}
\newcommand{\bigcell}[2]{\begin{tabular}{@{}#1@{}}#2\end{tabular}}
\usepackage{tabularx}
\newcolumntype{R}{>{\raggedleft\arraybackslash}X}
\newcolumntype{L}{>{\raggedright\arraybackslash}X}

\newtheorem{theorem}{Theorem}[section]

\newtheorem{assumption}{Assumption}

\begin{document}

\def\spacingset#1{\renewcommand{\baselinestretch}%
{#1}\small\normalsize} \spacingset{1}


\if1\blind
{
  \title{ A Joint Estimation Approach to Sparse Additive Ordinary Differential Equations}
  \author{Nan Zhang, Muye Nanshan
\hspace{.2cm}\\
    School of Data Science, Fudan University\\
    and \\
    Jiguo Cao \\
    Department of Statistics and Actuarial Science, Simon Fraser University}
  \date{}
  \maketitle
} \fi

\if0\blind
{
  \bigskip
  \bigskip
  \bigskip
  \begin{center}
    {\LARGE\bf A Joint Estimation Approach to Sparse Additive Ordinary Differential Equations}
\end{center}
  \medskip
} \fi

\begin{abstract}
Ordinary differential equations (ODEs) are widely used to characterize the dynamics of complex systems in real applications. In this article, we propose a novel joint estimation approach for generalized sparse additive ODEs where observations are allowed to be non-Gaussian. The new method is unified with existing collocation methods by considering the likelihood, ODE fidelity and sparse regularization simultaneously. We design a block coordinate descent algorithm for optimizing the non-convex and non-differentiable objective function. The global convergence of the algorithm is established. The simulation study and two applications demonstrate the superior performance of the proposed method in estimation and improved performance of identifying the sparse structure.
\end{abstract}

\noindent%

\spacingset{1.2} 

\section{Introduction}

Ordinary differential equation (ODE) models are popular to characterize the dynamics of complex systems in wide applications, such as infectious disease \citep{chen2008efficient,liang2010estimation}, gene regulation networks \citep{cao2008estimating,lu2011high} and brain connectivity \citep{zhang2015dynamic,wang2021network}. A general system of ODEs consisting of $p$ dynamic processes is given by
\begin{equation}\label{eq:general-ODE}
	\btheta^{'}(t{;}\bgamma)=F(\btheta(t{;}\bgamma),\bgamma),
\end{equation}
where $\btheta(t{;}\bgamma)=(\theta_1(t{;}\bgamma),\dots,\theta_p(t{;}\bgamma))^\top$, $F$ is usually a set of unknown functions describing the dependence between $\btheta(t{;}\bgamma)$ and its derivative $\btheta^{'}(t{;}\bgamma)$, and $\bgamma$ contains all the unknown parameters defining the system. Dynamic models in forms of ordinary differential equations \eqref{eq:general-ODE} associate the rate of change of processes $\btheta(t{;}\bgamma)$ with themselves. For instance, the gene regulation network uses the link function $F$ to quantify the regulatory effects of regulator genes on the expression change of a target gene \citep{wu2014sparse}. Neuronal state equations in a dynamic causal model \citep{friston2003dynamic} describe how instantaneous changes of the neuronal activities of system components are modulated jointly by the immediate states of other components.

ODE systems can be constructed by domain knowledge, such as the Newton's laws of motion in physics and the Lotka-Volterra equations for predator–prey biological system. Parameters of an ODE system have scientific interpretations. However, the values of these parameters are often unknown, and there is a strong need to estimate them from noisy observations measured from the dynamic system. Let $\mathcal{T}\subset\mathbb{R}$ be a compact interval and consider a common set of discrete time points $t_1,\dots, t_n\in\mathcal T$ for notation brevity. Denote by $y_{ij}$ the observation according to the $j$th latent process $\theta_j(t)$ at time $t=t_i$, $j=1,\dots,p$. To facilitate a flexible modeling, we assume the observations conditional on the latent processes follow an exponential family distribution, i.e., the conditional density function $f(y_{ij} {\mid}\, \theta_j(t_i))$ is given by
\begin{equation*}
	\exp\left\{\frac{y_{ij}\theta_j(t_i)-b(\theta_j(t_i))}{a(\phi)} + c(y_{ij},\phi)\right\},
\end{equation*}
where $a>0, b, c$ are known functions, $\phi$ is either known or considered as a nuisance parameter. The exponential family covers many continuous and discrete probability distributions, including Gaussian, Bernoulli, binomial, Poisson and negative binomial. See \cite{wood2017generalized} for a comprehensive review.

Parameter estimation for ODEs from noisy observations, also known as system identification, remains a challenging task in the statistical literature. Many methods have been developed including the nonlinear least squares \citep{biegler1986nonlinear}, the two-stage collocation methods \citep{varah1982spline,liang2008parameter,lu2011high,brunel2014parametric,wu2014sparse,dattner2015optimal,brunton2016discovering,chen2017network,dai2021kernel}, the parameter cascading methods \citep{ramsay2007parameter,cao2007parameter,qi2010asymptotic,nanshan2022dynamical}, and Bayesian methods \citep{huang2006hierarchical,zhang2017bayesian}. Based on their different treatment of the differential equations, we summarize them into three categories. The first approach requires the latent processes to follow the ODEs exactly. It searches for the optimal ODE parameters by evaluating the fitting between the data and the numerical solutions to the ODEs. This is named as the gold standard approach by \cite{chen2017network}. Since the numerical solution to the differential equations is sensitive to the values of ODE parameters, this approach suffers from numerical instability and intensive computation. The second approach is the two-stage collocation suggested by \cite{varah1982spline}. The latent processes and their derivatives are estimated via data smoothing procedures at the first stage, followed by treating the estimated derivative as a response variable and minimizing a least squares measure based on the differential equations with respect to the ODE parameters. Compared to the gold standard approach that latent processes must comply with the ODEs, the two-stage collocation uses the differential equations only as a discrepancy measure. The performance relies heavily on a satisfactory estimate of the derivatives, which can be challenging in nonparametric smoothing. Recently, \cite{dattner2015optimal}, \cite{chen2017network} and \cite{dai2021kernel} considered an integrated form of the ODEs to address this issue. The third approach is generalized profiling or parameter cascading \citep{ramsay2007parameter,cao2007parameter,nanshan2022dynamical}. It provides a nested optimization procedure to iteratively update the estimates of the latent processes regularized by their adherence to the differential equations. It has been shown to provide more efficient parameter estimation with theoretical guarantee \citep{qi2010asymptotic}.

Motivated by large-scale time-course data analysis, there have been many efforts devoted to high-dimensional ODEs under various model assumptions. For example, \citet{lu2011high} considered the high-dimensional linear ODEs to identify dynamic gene regulatory network, while \citet{zhang2015dynamic} studied a similar parametric model but allowed for two-way interactions. \citet{brunton2016discovering} proposed SINDy to identify the nonlinear dynamical system from a rich library of basis functions, which allows for more general interactions. \citet{henderson2014network}, \citet{wu2014sparse} and \citet{chen2017network} relaxed the parametric assumption by studying sparse additive ODEs. \citet{dai2021kernel} developed the kernel ODEs with functional ANOVA structure, which allows for pairwise interactions in a nonparametric fashion and gains greater flexibility. Regarding parameter estimation, all the above procedures for high-dimensional ODEs belong to the two-stage collocation category. Its popularity in parameter estimation for ODEs comes from several aspects. 
First, the two-stage methods have the so-called decoupled property \citep{liang2008parameter,lu2011high} that the estimation procedure for the $p$-dimensional ODE system can be decomposed into $p$ one-dimensional ODEs independently. The number of ODE parameters to be estimated grows exponentially with the number of differential equations. In particular, in an additive ODE model, the nonparametric components need to be approximated with basis expansions, which leads to even more model parameters. The decoupled property effectively mitigates the computational burden in the high-dimensional setting. 
Second, the two-stage collocation methods are convenient to incorporate sparsity-inducing penalties when we expect a sparse ODE structure for better model interpretation \citep{wu2014sparse,chen2017network}. Although the generalized profiling approach provides efficient ODE parameter estimation by improved latent process smoothing, its extension to high-dimensional ODEs is unclear, partially because of its nested optimization procedures. In summary, it is a distinctive challenge to simultaneously balance latent process smoothing, ODE parameter estimation, and variable selection for high-dimensional ODE models.

In this article, we propose a novel joint estimation approach for generalized sparse additive ODEs (JADE). To simultaneously accomplish the aforementioned tasks, we formulate an objective function consisting of a data fidelity associating the observed data with the latent processes, an ODE fidelity measuring the adherence of the latent processes to the differential equations, and a sparsity regularization controlling the model complexity. As a joint estimation strategy, JADE uses the ODE fidelity to regularize the estimation of latent processes, likewise the generalized profiling approach \citep{ramsay2007parameter}, and thus provides more accurate estimation in both the processes and their derivatives. For recovering the sparse structure of the ODE system, JADE approximates the additive functional components with finite basis expansions and naturally integrates group-wise sparse penalties to achieve the selection of individual functionals. Furthermore, under the regularized estimation framework, the objective function of JADE is non-convex and non-differentiable. To tackle this computational challenge, we design a block coordinate descent algorithm to solve the non-convex optimization problem with proper parameter tuning. By comparing with other existing methods via extensive numerical studies, we show that JADE not only provides more stable and accurate latent process smoothing and ODE parameter estimation, but also has good sparse structure recovery performance.

Our main contributions are summarized as follows. First, we allow the noisy observed data to be non-Gaussian, which covers a variety of continuous and discrete distributions for practical interest. Most existing literature on ODEs is based on the nonlinear least squares criterion \citep{miao2014generalized}, while the extension to likelihood-based criterion requires more involved computation. Second, we present a unified estimation framework under which JADE is a joint approach, whereas the two-stage collocation methods and the generalized profiling approach can be understood as conditional and profile estimation strategies, respectively. It also justifies the central role played by the ODE fidelity component in bridging the latent processes with the differential equations. Third, due to the non-convex nature of our objective function, the classical iteratively re-weighted least squares algorithm \citep{wood2017generalized} for generalized linear modeling performs poorly. Instead, we adopt a gradient-descent method for optimization and achieve great generalization performance. Fourth, we prove the convergence of our block coordinate descent algorithm to a stationary point of the objective function. The success of our algorithm hinges on the block-separable structure of the problem and an inexact line search instead of an exact Hessian calculation. Not only are the gradient-based iterations cheaper, but also stronger global convergence is possible.

The rest of this paper is organized as follows. We review existing methods and develop our JADE in Section~\ref{sec:jade}. Detailed computational procedure and convergence analysis are presented in Section~\ref{sec:bcd}. We investigate the numerical performance in Section~\ref{sec:simu} and illustrate with data examples in Section~\ref{sec:data}. Section~\ref{sec:conclude} provides some concluding remarks.

\section{Joint Estimation for Sparse Additive ODEs}\label{sec:jade}

In this section, we first briefly review existing approaches to ODE parameter estimation and then present a unified estimation framework for latent processes, ODE parameters and sparse ODE structure, under which our method is developed from the joint estimation perspective. It is thus dubbed joint estimation for generalized sparse additive ordinary differential equations (JADE).

Under the additive assumption, the ODE system \eqref{eq:general-ODE} takes the form
\begin{equation}\label{eq:additive-ODE}
	\theta^{'}_j(t) = \gamma_{j0} + \sum_{k=1}^pf_{jk}(\theta_k(t)),\  j= 1,\ldots,p.
\end{equation}
The additive components $f_{jk}$'s are usually approximated with basis expansions, for example, B-spline basis for numerical stability, such that 
\begin{equation}\label{eq:fjk_expansion}
	f_{jk}(\theta)= \bgamma_{jk}^\top \bphi(\theta)+\delta_{jk}(\theta),
\end{equation}
where $\theta$ denotes the value of a generic latent variable, $\bphi(\theta)=(\phi_1(\theta),\dots,\phi_L(\theta))^\top$ is the vector of basis functions, $\bgamma_{jk}\in \mathbb{R}^L$ is the vector of basis coefficients, and $\delta_{jk}$ is the residual function. Therefore, the additive ODE system \eqref{eq:additive-ODE} can be written as
\begin{equation}\label{eq:additive-ODE-bspl}
	\begin{aligned}
		\theta^{'}_j(t) 
		=\gamma_{j0} + \sum_{k=1}^p \bgamma_{jk}^\top \bphi(\theta_k(t))
		+\sum_{k=1}^p \delta_{jk}(\theta_k(t)),\ j= 1,\ldots,p.
	\end{aligned}
\end{equation}
Denote by $\bgamma_j=(\gamma_{j0}, \bgamma_{j1},\dots,\bgamma_{jp})$ the ODE parameters in the $j$th differential equation and let $\bgamma=(\bgamma_1,\dots,\bgamma_p)$ collect ODE parameters from all differential equations.

\subsection{Two-stage Collocation Methods}\label{sec:twostage}

Collocation with spline bases is first proposed by \cite{varah1982spline} for dynamic data fitting problems and has been successfully extended to parameter estimation and network reconstruction for high-dimensional ODE models \citep{liang2008parameter,lu2011high,brunel2014parametric,henderson2014network,wu2014sparse,dattner2015optimal,chen2017network,dai2021kernel}. It is a two-stage procedure in which the latent processes and their derivatives are first fitted by smoothing methods, followed by estimating the ODE parameters given the smoothed estimates. In particular, it solves the following optimization problems, for $j=1,\dots,p$,
\begin{align*}
		\wh \bgamma_j = \argmin_{\substack{\gamma_{j0}\in\mathbb R,\bgamma_{jk}\in\mathbb R^L}}\, \int_\mathcal{T} \bigg\{\frac{\rmd\wh\theta_j(t)}{\rmd t}-\gamma_{j0}-\sum_{k=1}^p \bgamma_{jk} \bphi(\wh\theta_k)\bigg\}^2 \,\rmd t + \sum_{k=1}^p \textrm{pen}_{\lambda_\gamma}(\|\bgamma_{jk}\|_2),
\end{align*}
with
\begin{equation*}
	\wh\theta_j(t) = \argmin_{\theta \in \mathcal H} -\frac{1}{n}\sum_{i=1}^n\{y_{ij}\theta(t_i)-b(\theta(t_i))\},
\end{equation*}
where $\textrm{pen}_{\lambda_\gamma}(\cdot)$ is a penalty function inducing sparse structure for the ODE system, $\mathcal H$ is a closed ball in a reproducing kernel Hilbert space, and $\wh\theta_j$ is known as the exponential family smoothing spline estimator \citep{wahba1995,gu2013smoothing}.

More involved two-stage collocation methods have been proposed for improvement. \citet{wu2014sparse} developed a five-step variable selection procedure for the sparse additive ODE model \eqref{eq:additive-ODE}, which refines the two-stage collocation approach with the standard group lasso and adaptive group lasso regularizations. However, considerable care is required in the smoothing step to ensure a satisfactory estimation of the latent process derivatives. \citet{dattner2015optimal} and \citet{chen2017network} further proposed a more robust ODE parameter estimation strategy for the second stage. Specifically, they used an integrated form of the differential equations to build up an objective function and successfully avoided estimating the derivatives. The two-stage collocation has an attractive decoupled property. Once the latent processes are estimated and fixed, the second stage procedure can be performed for each differential equation separately \citep{varah1982spline,wu2014sparse,chen2017network,dai2021kernel}. Computational efficiency together with theoretical guarantee leads to the prevalence of the two-stage collocation methods for parameter estimation and variable selection for high-dimensional ODE models.

\subsection{Generalized Profiling Procedure}\label{sec:gp}

\citet{ramsay2007parameter} proposed the generalized profiling approach for parameter estimation of ODEs. It is based on a generalized data smoothing strategy along with a generalization of profiled estimation. The ODE parameters $\{\bgamma_j\}_{j=1}^p$ are of primary interest, while the latent processes $\{\theta_j(t)\}_{j=1}^p$ are nuisance. In contrast to the two-stage collocation methods, the generalized profiling approach involves a nested optimization procedure. During the inner optimization of the profiling procedure, an intermediate estimate of the latent processes $\wh\btheta(t,\bgamma)=(\wh\theta_1(t,\bgamma),\dots,\wh\theta_p(t,\bgamma))$ is obtained by minimizing a combination of data and ODE fidelity criteria
\begin{align}\label{eq:gp-inner}
	-\frac{1}{n}\sum_{j=1}^p\sum_{i=1}^n \left\{y_{ji}{\theta_j(t_i)}-b(\theta_j(t_i))\right\} \nonumber
	+ \lambda_\theta \sum_{j=1}^p\int_\mathcal{T} \bigg\{\frac{\rmd\theta_j(t)}{\rmd t}-\gamma_{j0}-\sum_{k=1}^p \bgamma_{jk}^\top \bphi(\theta_k(t))\bigg\}^2 \rmd t,
\end{align}
where a larger value of the tuning parameter $\lambda_\theta>0$ encourages more adherence of the latent processes to the differential equation system. A data fitting criterion, such as log-likelihood, is then optimized with respect to the ODE parameters $\bgamma$ by holding $\wh\btheta(t,\bgamma)$ as a function of $\bgamma$, that is, 
\begin{equation*}
	\wh\bgamma = \argmin_{\bgamma} -\frac{1}{n}\sum_{j=1}^p\sum_{i=1}^n \big\{y_{ji}{\wh\theta_j(t_i{;}\bgamma)}-b(\wh\theta_j(t_i{;}\bgamma))\big\}.
\end{equation*}
The generalized profiling approach proceeds with solving inner and outer optimizations iteratively with a non-decreasing sequence of $\lambda_\theta$. Although the generalized profiling approach provides a more efficient estimation strategy for small-scale ODEs, its extension to the high-dimensional ODEs remains challenging. Unlike the two-stage collocation methods, the optimization procedure of the generalized profiling cannot be decoupled for each latent process or differential equation. It is computationally demanding to handle a large number of primary and nuisance parameters. Moreover, when a sparsity-inducing penalty is introduced in the outer optimization to control the model complexity, the objective function is not convex such that the Gauss-Newton method used by \cite{ramsay2007parameter} is no longer applicable.

\subsection{Joint Estimation Approach}

To facilitate latent process and ODE parameter estimation together with sparse structure identification for sparse additive ODEs, we formulate a unified estimation framework and build our JADE procedure. Combining the ingredients of the two-stage collocation and generalized profiling methods, we define an objective function consisting of three components: a data fidelity associating the observations with the latent processes, an ODE fidelity measuring the adherence of latent processes to the differential equations, and a sparsity regularization controlling the model complexity. Specifically, JADE aims to simultaneously estimate the latent processes $\btheta(t)$ and ODE parameters $\bgamma$ by minimizing
\begin{equation}\label{eq:jointest}
	\begin{split}
		Q(\btheta(t),\bgamma)&=
		-\frac{1}{n}\sum_{j=1}^p\sum_{i=1}^n \left\{y_{ji}{\theta_j(t_i)}-b(\theta_j(t_i))\right\} \\
		&\qquad+ \lambda_\theta \sum_{j=1}^p\int_\mathcal{T} \bigg\{\frac{\rmd\theta_j(t)}{\rmd t}-\gamma_{j0}-\sum_{k=1}^p \bgamma_{jk}^\top \, \bphi(\theta_k(t))\bigg\}^2 \rmd t \\
		&\qquad+ \sum_{k=1}^p \textrm{pen}_{\lambda_\gamma}(\|\bgamma_{jk}\|_2),
	\end{split}
\end{equation}
where $\lambda_\theta, \lambda_\gamma>0$ are tuning parameters balancing the ODE fidelity and model complexity, respectively. The penalty function $\textrm{pen}_{\lambda_\gamma}(\cdot)$, for example, the group lasso \citep{yuan2006model} or its variants, introduces group-wise sparsity by forcing all elements in $\bgamma_{jk}$ to be either zero or nonzero and thus allows for identifying significant additive components in the ODE system. 

Our JADE procedure views both the latent processes and ODE parameters of primary interest. Observe that the latent processes $\btheta(t)$ appear in the first two components in \eqref{eq:jointest}, while the ODE parameters $\bgamma$ are associated with the last two components. It indicates that the ODE fidelity plays a central role in bridging the latent processes with the ODE structure. When latent processes are approximated with basis expansions, the basis coefficients and ODE parameters naturally fit in a block optimization architecture. A closer look at $Q(\btheta(t),\bgamma)$ in \eqref{eq:jointest} reveals more insight. On the one hand, for the latent processes or equivalently their basis coefficients, the likelihood term in $Q$ is convex, but the ODE fidelity is non-convex due to the use of basis $\bphi$. On the other hand, the expression of differential equations in the second component of \eqref{eq:jointest} is linear in $\bgamma_j$'s. Thus, optimizing $Q$ with respect to the ODE parameters is essentially solving group-regularized least squares. Computational details of JADE are presented in Section~\ref{sec:bcd}.

We conclude this section by comparing with other estimation strategies under the unified estimation framework. The two-stage collocation methods first obtain $\wh\btheta(t)$ from the data fidelity component and then minimize the latter two components in \eqref{eq:jointest} for an estimate of $\bgamma$ by conditioning on $\wh\btheta(t)$. The estimated latent processes are only based on observations but are not required to satisfy the differential equations. On the other hand, the generalized profiling approach does not consider sparsity regularization. It treats $\btheta$ as a nuisance parameter, where the inner optimization deduces the dependence of $\wh\btheta(t, \bgamma)$ on $\bgamma$ by minimizing the first two components in \eqref{eq:jointest}, followed by profiling out $\wh\btheta(t, \bgamma)$ and optimizing the data fidelity with respect to $\bgamma$ iteratively. In summary, from the estimation perspective, the two-stage collocation methods follow a conditional strategy, and the generalized profiling approach is in a profiling fashion.

\section{Block Coordinate Descent Algorithm}\label{sec:bcd}

We develop a block coordinate descent algorithm for our JADE procedure and establish its global convergence. Like collocation based methods, we use a finite basis to approximate the $j$th latent process with $\wt\theta_j(t) = \bc_j^\top\bpsi(t)$, where $\bc_j \in \mathbb{R}^M$, $\bpsi(t)=(\psi_1(t),\ldots,\psi_M(t))^\top$ and $M$ is the number of basis functions. We assume $M$ to be the same for $j=1,\dots,p$ without loss of generality. Now the objective function of JADE in \eqref{eq:jointest} can be expressed as a function of basis coefficients of latent processes and ODE additive components, denoted by $\bc=(\bc_1,\dots,\bc_p)$ and $\bgamma=(\bgamma_1,\dots,\bgamma_p)$, respectively. The parameters to be optimized naturally form $2p$ blocks, which prompts us to consider a block coordinate descent algorithm. 

With some abuse of notation, we use $Q(\bc, \bgamma)$ to denote the quantity in \eqref{eq:jointest} with $\btheta(t)$ being replaced by their basis expansions. We will not distinguish them in the following discussion since the definition should be clear in context. According to \eqref{eq:jointest}, we write 
$$
Q(\bc,\bgamma)=S(\bc,\bgamma)+R(\bgamma),
$$
where
\begin{align*}
	S(\bc,\bgamma)&=-\frac{1}{n}\sum_{j=1}^p\sum_{i=1}^n \left\{y_{ji}{\bc_j^\top\bpsi(t_i)}-b(\bc_j^\top\bpsi(t_i))\right\} \\
	 &\qquad\qquad+ \lambda_\theta \sum_{j=1}^p\int_\mathcal{T} \bigg\{\bc_j^\top\frac{\rmd\bpsi(t)}{\rmd t}-\gamma_{j0}-\sum_{k=1}^p \bgamma_{jk}^\top \, \bphi(\bc_k^\top\bpsi(t))\bigg\}^2\rmd t
\end{align*}
adds up the likelihood and the ODE fidelity terms, and
$R(\bgamma)=\sum_{k=1}^p \textrm{pen}_{\lambda_\gamma}(\|\bgamma_{jk}\|_2)$
regularizes the sparsity of $\bgamma$. We have two key observations for the above decomposition. First, $S(\bc,\bgamma)$ is a continuously differentiable function of all blocks of parameters. Particularly, it is a block-separable quadratic function of $\bgamma_j$'s, but is non-convex in $\bc_j$'s due to the additive assumption of the ODEs. Second, $R(\bgamma)$ encourages group-wise sparsity for $\bgamma$ and thus has a block-separable structure, i.e., $R(\bgamma) = \sum_{j=1}^p R_j(\bgamma_j)$. 

Sections~\ref{sec:bcd_theta} and \ref{sec:bcd_gamma} describe the updating rules for the latent processes and the additive components in the ODE system, respectively. We then address identifiability and parameter tuning issues and analyze the global convergence of the proposed method.

\subsection{Update the Latent Processes}\label{sec:bcd_theta}

Recall that for $j=1,\dots,p$, the $j$th latent process $\theta_j(t)$ is approximated by basis expansion $\bc_j^\top\bpsi(t)$, and the JADE objective function $Q(\bc,\bgamma)$ depends on $\bc_j$ only through $S(\bc,\bgamma)$. Let $\nabla_{\bc_j} S(\bc, \bgamma)\in\mathbb R^M$ be the gradient of $S(\bc,\bgamma)$ with respect to $\bc_j$. We use it to build a quadratic approximation to $S(\bc, \bgamma)$ with respect to $\bc_j$ and generate an improving direction. More precisely, we choose a positive definite matrix $\bH_{\bc_j} \in \mathbb{R}^{M\times M}$ and move $\bc_j$ along the direction $\bd_{\bc_j}(\bc,\bgamma)$, where 
\begin{equation}\label{eq:dcj}
	\begin{aligned}
		\bd_{\bc_j}(\bc,\bgamma) &=\argmin_{\bd \in \mathbb{R}^M} \big\{ (\nabla_{\bc_j} S(\bc, \bgamma))^\top \bd + \frac{1}{2} \bd^\top \bH_{\bc_j} \bd\big\}
		= -\bH_{\bc_j}^{-1}\,(\nabla_{\bc_j} S(\bc, \bgamma)).
	\end{aligned}
\end{equation}
The above first equality implies that $\bd_{\bc_j}(\bc,\bgamma)$ minimizes the quadratic perturbation of $S(\bc,\bgamma)$ caused by the change of $\bc_j$. Because $S(\bc,\bgamma)$ is continuously differentiable in $\bc_j$, a typical choice of $\bH_{\bc_j}$ is the Hessian matrix $\nabla^2_{\bc_j} S(\bc, \bgamma)$. In this case, the JADE algorithm is equivalent to the Newton-type strategy to minimize $S(\bc, \bgamma)$. However, it can be expensive to compute all Hessian matrices for blocks $\bc_j$'s, and the Hessian matrices are not guaranteed to be positive definite because $S(\bc, \bgamma)$ is non-convex in $\bc$. Based on the above considerations, we follow \cite{tseng2009coordinate} and choose a diagonal Hessian approximation $\bH_{\bc_j}$, whose diagonal elements are the same as those of $\nabla^2_{\bc_j} S(\bc, \bgamma)$ thresholded within a positive interval $[\underline{\mu},\overline{\mu}]$ where $0<\underline{\mu}\leq\overline{\mu}$. Such a choice satisfies the positive definite requirement and allows for cheaper iterations than the exact Hessian matrices.

Once the descent direction $\bd_{\bc_j}(\bc,\bgamma)$ is obtained, we implement a line search with the step size $\alpha_j>0$ selected by the Armijo rule \citep{burke1985descent, nocedal2006numerical,fletcher2013practical}, see detail in Appendix~\ref{sec:Armijo}. Finally, the basis coefficient of the $j$th latent process is updated by $\bc_j + \alpha_j \bd_{\bc_j}(\bc,\bgamma)$. 

Before moving forward to update the ODE system, we remark that in \eqref{eq:fjk_expansion} we use basis $\bphi(\theta)$ to expand the additive components $f_{jk}$'s. It is common to assume $\bphi$ is defined on a compact support $\mathcal{I}$. However, when $\wt\theta_j(t) = \bc_j^\top\bpsi(t)$ is updated for approximating the $j$th latent process, its range may exceed $\mathcal{I}$. To resolve this issue, we exploit a smooth and monotone transformation map $\sigma: \mathbb R \to \mathcal I$ to construct a composite basis $\bphi(\sigma(\wt\theta_j))$. Choices of $\sigma(\theta)$ can be the sigmoid function class, including the logistic function $(1+e^{-\theta})^{-1}$, the hyperbolic tangent $\tanh(\theta)$ and the arctangent function $\arctan(\theta)$. For notation brevity, we still use $\bphi(\cdot)$ as the basis function composed with transformation $\sigma(\cdot)$ without causing confusion.

\subsection{Update the Additive Components in the ODE System}\label{sec:bcd_gamma} 

Consider updating the ODE parameters in the $j$th differential equation $\bgamma_j=(\gamma_{j0},\bgamma_{j1},\dots,\bgamma_{jp})$, where $\gamma_{j0}$ is the constant intercept and $\bgamma_{jk}$ is the vector of basis coefficients when $f_{jk}$ is expanded with the basis functions $\bphi$, $k=1,\dots,p$. Given the latent process estimates $\{\wt\theta_j(t): 1\leq j\leq p\}$ and the ODE parameters in other differential equations, minimizing $Q(\bc,\bgamma)$ with respect to $\bgamma_j$ amounts to solving
\begin{equation}\label{eq:bcd_gamma}
	\begin{aligned}
		\argmin_{\bgamma_j\in\mathbb R^{pL+1}} \int_\mathcal{T} \bigg\{\frac{\rmd\wt\theta_j(t)}{\rmd t}-\gamma_{j0}-\sum_{k=1}^p\bgamma_{jk}^\top \, \bphi(\wt\theta_k(t))\bigg\}^2 \rmd t  +\sum_{k=1}^p\textrm{pen}_{\lambda_\gamma}(\|\bgamma_{jk}\|_2),
	\end{aligned}
\end{equation}
which coincides with a penalized regression problem with group selection. Examples of $\textrm{pen}_{\lambda_\gamma}(\cdot)$ include group lasso, group SCAD and group MCP. See \cite{huang2012selective} and \cite{breheny2015group} for a comprehensive review of efficient algorithms. 

In our JADE algorithm, we choose the adaptive group lasso penalty $\textrm{pen}_{\lambda_\gamma}(\|\bgamma_{jk}\|_2) = \lambda_\gamma\, w_{jk} \|\bgamma_{jk}\|_2$. It has been shown to achieve both estimation efficiency and selection consistency \citep{wang2008note}. Let $\{\wt\bgamma_{jk}, 1\leq k\leq p\}$ be the regular group lasso estimates. We follow \cite{zou2006adaptive} to set the weight $w_{jk}=\|\wt\bgamma_{jk}\|_2^{-\nu}$ if $\|\wt\bgamma_{jk}\|_2>0$, where $\nu$ is a positive number, otherwise $w_{jk}=\infty$. A remarkable feature of updating the ODE parameters is that they can be performed in parallel due to the block separability of the ODE fidelity and group-wise sparse penalty in terms of $\bgamma_j$, $j=1,\dots,p$. Finally, the additive components in the ODE system are updated by implementing adaptive group lasso regressions in parallel. 
To summarize, we describe the updating rules of JADE in Algorithm~\ref{algo:jade}.
\begin{algorithm}[h!]
	\begin{enumerate}
		\item Initialize the basis coefficients $\bc^{0}$ and $\bgamma^{0}$ for the latent processes and additive components in the ODE system.
		
		\item At the $r$th step where $r\geq 1$, 
		\begin{enumerate}
			\item {\bf Update the the $j$th latent process:} obtain $\bd_{\bc_j}(\bc^r,\bgamma^r)$ as the descent direction and choose a step size $\alpha_j^{r}>0$; set $\bc_j^{r+1}\leftarrow\bc_j^{r}+\alpha_j^{r}\bd_{\bc_j}(\bc^r,\bgamma^r)$;
			
			\item {\bf Update the $j$th differential equation:} obtain $\bgamma_j^{r+1}$ via an adaptive group lasso regression.
			
		\end{enumerate}
		\item Repeat the above step until convergence or stop when the maximum number of iterations is reached.
	\end{enumerate}  
	\caption{Block coordinate descent algorithm for the JADE procedure.}
	\label{algo:jade}
\end{algorithm}

\subsection{Identifiability}

Because of the additive assumption of the ODE system, the ODE parameters are not fully identifiable. One can always add a constant to an additive component $f_{jk}$ in \eqref{eq:additive-ODE} and subtract it from another, and the model remains the same. It is related to the collinearity of the basis functions. To address this issue, we require that each additive component has the mean zero in the sense of its basis representation, that is, $\sum_{i=1}^{m} \bgamma_{jk}^\top\, \bphi(\theta_k(t_i))= 0$ for $1\leq j,k\leq p$, where $t_1,\dots, t_m\in\mathcal T$. Once the ODE parameters $\bgamma_j$ is updated, we subtract $\sum_{i=1}^{m} \bgamma_{jk}^\top\, \bphi(\theta_k(t_i))$ from the $k$th additive component and shift the intercept $\gamma_{j0}$ to $\gamma_{j0}+\sum_{k=1}^p \sum_{i=1}^{m} \bgamma_{jk}^\top\, \bphi(\theta_k(t_i))$.

\subsection{Tuning Parameter Selection}\label{sec:tuning}

Two tuning parameters appear in our JADE procedure.
The first one is $\lambda_\theta$ controlling the adherence of the latent processes to the ODE system. Similar to the generalized profiling approach \citep{ramsay2007parameter}, we expect a large $\lambda_\theta$ such that the ODEs are satisfied or approximately so. As pointed by \citet{qi2010asymptotic} and \citet{carey2021fast}, there exist many local optima when $\lambda_\theta$ is small, and $\lambda_\theta$ should be gradually increased together with monitoring the performance of the parameter estimates. However, our numerical experiments show that JADE is not sensitive to $\lambda_\theta$ as long as $\lambda_\theta$ is not too small, and we thereby set it as a fixed large constant. A numerical justification is provided in Section~\ref{sec:insenitive_lamtheta}.

The second tuning parameter $\lambda_\gamma$ controls the amount of sparse regularization on the ODE parameters. We propose to search for optimal $\lambda_\gamma$ by minimizing the following criterion
\begin{equation}\label{eq:bic-ode}
	\begin{aligned}
		\sum_{j=1}^p \log\bigg[ \int_\mathcal{T} \bigg\{\frac{\rmd\wh\theta_j(t)}{\rmd t}-\wh\gamma_{j0} -\sum_{k=1}^p\wh\bgamma_{jk}^\top \, \bphi(\wh\theta_k(t))\bigg\}^2 \,\rmd t \bigg] 
		 + \sum_{j=1}^p \mathrm{nz}(\wh\bgamma_{j}) \frac{\log(n)}{n},
	\end{aligned}
\end{equation}
where $\mathrm{nz}(\wh\bgamma_{j})$ is the number of non-zero elements in the ODE parameter $\wh\bgamma_{j}$. It shares a similar spirit of the Bayesian information criterion commonly used in the ODE parameter estimation literature \citep{wu2014sparse, chen2017network}. The first ODE fidelity term can be understood as likelihood measuring how compatible the estimated ODE parameters are given the latent process estimates, while the remaining term in \eqref{eq:bic-ode} penalizes the model complexity.

\subsection{Global Convergence Analysis}\label{sec:conv}

In our JADE algorithm, the latent processes are updated by a gradient descent method with the step size selected by the Armijo rule. We next show that the updating rule for the ODE parameters $\bgamma_j$ in Section~\ref{sec:bcd_gamma} is equivalent to a gradient descent method with the step size chosen by the Armijo rule. 

Recall that the penalty function $R(\bgamma)$ is block-separable such that $R(\bgamma) = \sum_{j=1}^p R_j(\bgamma_j)$. We define the gradient direction for minimizing $Q(\bc,\bgamma)$ with respect to $\bgamma_j$ as
\begin{equation}\label{eq:dgamj}
	\begin{aligned}
		\bd_{\bgamma_j}(\bc,\bgamma) =\argmin_{\bd \in \mathbb{R}^{pL+1}} \bigg\{ (\nabla_{\bgamma_j} S)^\top \bd + \frac{1}{2} \bd^\top (\nabla^2_{\bgamma_j} S) \bd 
		+ R_j(\bgamma_j+\bd) \bigg\},
	\end{aligned}
\end{equation}
where $\nabla_{\bgamma_j} S$ and $\nabla^2_{\bgamma_j} S$ are the gradient and Hessian of $S(\bc,\bgamma)$ with respect to $\bgamma_j$. Since $S(\bc,\bgamma)$ is quadratic in each block vector $\{\bgamma_j: 1\leq j\leq p\}$, the group penalized solution in \eqref{eq:bcd_gamma} amounts to updating $\bgamma_j$ to $\bgamma_j+\bd_{\bgamma_j}(\bc,\bgamma)$ while holding fixed other block vectors.
With a slight abuse of notation, the above gradient update leads to a change of the JADE objective function value given by
\begin{equation*}
	\begin{aligned}
		Q(\bc, \bgamma_{-j}, &\bgamma_j+\bd_{\bgamma_j}) - Q(\bc, \bgamma)
		=(\nabla_{\bgamma_j} S)^\top \bd_{\bgamma_j} + \frac{1}{2}\bd_{\bgamma_j}^\top (\nabla^2_{\bgamma_j} S)\, \bd_{\bgamma_j}
		 + R_j(\bgamma_j + \bd_{\bgamma_j}) - R_j(\bgamma_j).
	\end{aligned}
\end{equation*}
By the definition of $\bd_{\bgamma_j}(\bc,\bgamma)$ in \eqref{eq:dgamj}, the right-hand side of the above display is always non-positive, and thus we can set the step size to be constant one, which satisfies the Armijo rule \eqref{eq:armijo} in the Appendix.

To sum up, our JADE algorithm can be understood as a block coordinate gradient descent method with step sizes chosen by the Armijo rule. We need the following assumption for convergence analysis.
\begin{assumption}\label{ass:eigen}
	The eigenvalues of the matrices $\bH_{\bc_j}$ and the Hessians $\nabla^2_{\bgamma_j} S$ are bounded away from zero and infinity uniformly over $j=1,\dots,p$.
\end{assumption}
As discussed in Section~\ref{sec:bcd_theta}, the eigenvalues of $\bH_{\bc_j}$'s are bounded by the thresholding construction. The Hessians $\nabla^2_{\bgamma_j} S$, $j=1,\dots,p$, admit the following explicit expression
\begin{equation}\label{eq:hessian-matrix}
2\lambda_\theta \int_{\mathcal{T}} \overline{\bphi}(\btheta(t))\,\overline{\bphi}(\btheta(t))^\top\, \rmd t, 
\end{equation}
where $\overline{\bphi}(\btheta(t))$ is the augmented basis function vector concatenating the B-spline basis $\bphi(\theta_1(t)),\ldots,\bphi(\theta_p(t))$. 
The second part of Assumption~\ref{ass:eigen} is essentially made for the basis function $\bphi(\cdot)$, which is also adopted by \citet{chen2017network} to ensure identifiability. In fact, the eigenvalues of \eqref{eq:hessian-matrix} are upper bounded because of the boundedness of B-spline basis and the compactness of $\mathcal{T}$. When any two of the latent processes are not too close to each other over $\mathcal T$, the eigenvalues of \eqref{eq:hessian-matrix} can be determined by the diagonal blocks $\int_{\mathcal{T}} \bphi(\theta_k(t))\,\bphi(\theta_k(t))^\top\, \rmd t $, $k=1,\dots,p$, which are positive definite matrices due to the linear independence of B-splines \citep{de1978practical}. Finally, the global convergence result follows from Theorem~1 of \citet{tseng2009coordinate}.
\begin{theorem}
	Let $\{(\bc^r,\bgamma^r)\}$ be the sequence generated by the JADE Algorithm~\ref{algo:jade} under Assumption~\ref{ass:eigen}. If the step sizes $\{\alpha_j^r: 1\leq j\leq p\}$ are chosen by the Armijo rule with $\inf_{1\leq j\leq p} \alpha_{{\rm init},j}>0$ and $\sup_{j,r} \alpha_j^r<\infty$, then every cluster point of the sequence $\{(\bc^r, \bgamma^r)\}$ is a stationary point of the objective function $Q(\bc, \bgamma)$.
\end{theorem}

\section{Simulation Studies}\label{sec:simu}

In this section, we compare the performance of JADE procedure and two of the two-stage collocation methods, GRADE \citep{chen2017network} and SA-ODE \citep{wu2014sparse}, on sparse additive ODEs. 

\subsection{Set-up}\label{sec:setup}
We consider the example from \citet{chen2017network} where the number of latent processes $p=10$. The latent processes $\{\theta_j(t): 1\leq j\leq p, t\in\mathcal{T}=[0,20]\}$ satisfy the ODE system:
\begin{equation}\label{eq:ode_add_gen}
	\begin{cases}
		\theta'_{2k-1}(t) = \beta_{2k-1,0} + \bbeta^\top_{2k-1,2k-1}\,\bh(\theta_{2k-1}(t)) + \bbeta^\top_{2k-1,2k}\,\bh(\theta_{2k}(t)) \\
		\theta'_{2k}(t) = \beta_{2k,0} + \bbeta^\top_{2k,2k-1}\,\bh(\theta_{2k-1}(t)) + \bbeta^\top_{2k,2k}\,\bh(\theta_{2k}(t))
	\end{cases},
\end{equation}
where $k= 1,\ldots,5$, and $\bh(\theta) = (\theta, \theta^2, \theta^3)^\top$ is the cubic monomial basis. Among all $\bbeta_{jk}$'s where $1\leq j,k\leq 10$, eight of them are nonzero, while the rest ninety-two components are zero. Detailed specification of \eqref{eq:ode_add_gen} is presented in Appendix~\ref{sec:ode_appendix}. The system \eqref{eq:ode_add_gen} can be numerically solved using R package \verb|deSolve| given initial conditions.

Given the latent processes, we generate samples from Gaussian, Poisson and Bernoulli distributions, respectively. Let $t_1,\ldots,t_n$ be equally spaced over $\mathcal{T}$ with various sample sizes $n=40, 100$, and 200. For Gaussian distribution, $y_{ij}$ is drawn from $\mathcal{N}(\theta_j(t_i), \sigma^2)$ with known variance $\sigma^2$ at different levels. For Poisson distribution, we allow for $10$ replicated observations at each time point from $\textrm{Poisson}(\lambda_j(t_i))$, where the intensity $\lambda_j(t_i)=\exp(\theta_j(t_i))$. For Bernoulli distribution, we allow for $40$ replicated observations at each time point with the probability of success being  $p_j(t_i)=\exp\{\theta_j(t_i)\}/(1+\exp\{\theta_j(t_i)\})\}$.

In the JADE procedure, we initialize with the smoothing spline estimates for the latent processes, using the \verb|ssanova| suite for Gaussian observations and the \verb|gssanova| suite for Poisson or Bernoulli observations in R package \verb|gss|. When estimating the additive ODE components, we choose a cubic spline basis with four knots and the logistic transformation function $\sigma(\theta)=(1+e^{-\theta})^{-1}$. The block coordinate descent algorithm updates the parameters for a maximum of 4 iterations. GRADE and SA-ODE are implemented under their default configurations.

We evaluate the performance of different methods on two aspects. The first is the estimation performance for the latent processes and the additive ODE components. We use the following two averaged mean squared errors (MSE) to measure the estimation error of the latent processes and their derivatives, respectively,
\begin{align*}
	\mathrm{MSE}(\wh\btheta(t)) &= \frac{1}{p} \sum_{j=1}^{p} \int_{\mathcal{T}} \left\{ \wh\theta_j(t) - \theta_j(t) \right\}^2 \rmd t, \\
	\mathrm{MSE}(\wh\btheta'(t)) &= \frac{1}{p} \sum_{j=1}^{p} \int_{\mathcal{T}} \left\{ \wh\theta'_j(t) - \theta^{'}_j(t) \right\}^2 \rmd t.
\end{align*}
To better characterize the estimation in the additive ODE components, we assess the estimation error on the active set $\mathcal A=\{f_{jk}:f_{jk}\not\equiv 0\}$ and inactive set $\mathcal A^c$ separately, where the universe set consists of all additive components. Specifically, we consider
\begin{align*}
	&\mathrm{MSE}_{\textrm{active}}(\wh f)= \frac{1}{\lvert\mathcal A\rvert} \sum_{f_{jk}\in \mathcal A} \int_{\mathcal{R}_k} \left\{\wh f_{jk}(\theta) - f_{jk}(\theta)\right\}^2 \rmd \theta,\\
	&\mathrm{MSE}_{\textrm{inactive}}(\wh f)= \frac{1}{\lvert\mathcal A^c\rvert} \sum_{f_{jk}\in \mathcal A^c} \int_{\mathcal{R}_k} \left\{\wh f_{jk}(\theta) - f_{jk}(\theta)\right\}^2 \rmd \theta,
\end{align*}
where $\lvert\cdot\rvert$ calculates the cardinality of a set, and $\mathcal{R}_k$ is the range of $\theta_k(t)$ on $\mathcal{T}$. The second aspect is the network discovery. True-positive rates and false-positive rates are used to describe the accuracy of identifying nonzero additive ODE components. We repeat each experiment for 100 times and present averaged performance evaluations in Tables~\ref{tab:res_gaussian}, \ref{tab:res_poisson} and \ref{tab:res_binomial}. 

\subsection{Results}
\begin{table*}[htbp]
	\centering
	\caption{Gaussian observation. Comparison of methods in latent process fitting, ODE additive components estimation and network discovery. Standard errors are displayed in the parentheses under the evaluation scores.}
	\ra{1.2}
	\tiny
	\begin{tabularx}{\textwidth}{@{}rrrRRRRRR@{}}
		\toprule
		$n$ & SNR & Method 
		& \multicolumn{2}{c}{Latent Process} & \multicolumn{2}{c}{ODE Additive Component} & \multicolumn{2}{c}{Network Discovery} \\ 
		\midrule
		&  &  & \bigcell{c}{ $\textrm{MSE}(\wh\btheta)$ \\[-0pt] ($\times10^{-2}$)}   & $\textrm{MSE}(\wh \btheta' )$ & $\textrm{MSE}_{\textrm{active}}(\wh f)$ & $\textrm{MSE}_{\textrm{inactive}}(\wh f)$ & TP\% & FP\%  \\
		\cmidrule{4-9}
		
		\multirow{3}{*}{ 200 } & \multirow{3}{*}{ 25 } & JADE & \bigcell{r}{ \textbf{0.030} \\[-2pt]  (0.007) } & \bigcell{r}{ \textbf{0.005} \\[-2pt]  (0.001) } & \bigcell{r}{ \textbf{0.033} \\[-2pt]  (0.024) } & \bigcell{r}{ 0.007 \\[-2pt]  (0.007) } & \bigcell{r}{ \textbf{100} \\[-2pt]  (0.0) } & \bigcell{r}{ 20.4 \\[-2pt]  (7.0) } \\
		&   & GRADE & \multirow{2}{*}{ \bigcell{r}{ 0.053 \\[-2pt]  (0.006) } } & \multirow{2}{*}{ \bigcell{r}{ 0.015 \\[-2pt]  (0.002) } } & \bigcell{r}{ 0.056 \\[-2pt]  (0.007) } & \bigcell{r}{ \textbf{0.001} \\[-2pt]  (0.000) } & \bigcell{r}{ \textbf{100} \\[-2pt]  (0.0) } & \bigcell{r}{ 35.9 \\[-2pt]  (3.8) } \\
		&   & SA-ODE &   &   & \bigcell{r}{ 0.060 \\[-2pt]  (0.052) } & \bigcell{r}{ 0.005 \\[-2pt]  (0.007) } & \bigcell{r}{ \textbf{100} \\[-2pt]  (0.0) } & \bigcell{r}{ \textbf{8.6} \\[-2pt]  (3.6) } \\
		\cmidrule{1-9}
		
		\multirow{3}{*}{ 200 } & \multirow{3}{*}{ 10 } & JADE & \bigcell{r}{ \textbf{0.194} \\[-2pt]  (0.030) } & \bigcell{r}{ \textbf{0.023} \\[-2pt]  (0.004) } & \bigcell{r}{ 0.147 \\[-2pt]  (0.091) } & \bigcell{r}{ 0.073 \\[-2pt]  (0.055) } & \bigcell{r}{ 99.7 \\[-2pt]  (2.0) } & \bigcell{r}{ 30.3 \\[-2pt]  (7.1) } \\
		&   & GRADE & \multirow{2}{*}{ \bigcell{r}{ 0.258 \\[-2pt]  (0.026) } } & \multirow{2}{*}{ \bigcell{r}{ 0.040 \\[-2pt]  (0.006) } } & \bigcell{r}{ \textbf{0.124} \\[-2pt]  (0.027) } & \bigcell{r}{ \textbf{0.001} \\[-2pt]  (0.001) } & \bigcell{r}{ 98.8 \\[-2pt]  (3.8) } & \bigcell{r}{ 27.6 \\[-2pt]  (4.9) } \\
		&   & SA-ODE &   &   & \bigcell{r}{ 0.274 \\[-2pt]  (0.220) } & \bigcell{r}{ 0.149 \\[-2pt]  (0.212) } & \bigcell{r}{ \textbf{100} \\[-2pt]  (0.0) } & \bigcell{r}{ \textbf{22.2} \\[-2pt]  (5.8) } \\
		\cmidrule{1-9}
		
		\multirow{3}{*}{ 200 } & \multirow{3}{*}{ 4 } & JADE & \bigcell{r}{ \textbf{1.102} \\[-2pt]  (0.128) } & \bigcell{r}{ \textbf{0.084} \\[-2pt]  (0.010) } & \bigcell{r}{ 0.355 \\[-2pt]  (0.117) } & \bigcell{r}{ 0.279 \\[-2pt]  (0.199) } & \bigcell{r}{ \textbf{100} \\[-2pt]  (0.0) } & \bigcell{r}{ 46.4 \\[-2pt]  (10.9) } \\
		&   & GRADE & \multirow{2}{*}{ \bigcell{r}{ 1.123 \\[-2pt]  (0.134) } } & \multirow{2}{*}{ \bigcell{r}{ 0.094 \\[-2pt]  (0.016) } } & \bigcell{r}{ \textbf{0.316} \\[-2pt]  (0.054) } & \bigcell{r}{ \textbf{0.004} \\[-2pt]  (0.006) } & \bigcell{r}{ 85.6 \\[-2pt]  (8.4) } & \bigcell{r}{ \textbf{16.5} \\[-2pt]  (3.1) } \\
		&   & SA-ODE &   &   & \bigcell{r}{ 0.566 \\[-2pt]  (0.259) } & \bigcell{r}{ 1.791 \\[-2pt]  (2.461) } & \bigcell{r}{ 98.5 \\[-2pt]  (4.1) } & \bigcell{r}{ 36.8 \\[-2pt]  (7.6) } \\
		\cmidrule{1-9}
		
		\multirow{3}{*}{ 100 } & \multirow{3}{*}{ 25 } & JADE & \bigcell{r}{ \textbf{0.049} \\[-2pt]  (0.010) } & \bigcell{r}{ \textbf{0.007} \\[-2pt]  (0.001) } & \bigcell{r}{ \textbf{0.042} \\[-2pt]  (0.021) } & \bigcell{r}{ 0.006 \\[-2pt]  (0.010) } & \bigcell{r}{ \textbf{100} \\[-2pt]  (0.0) } & \bigcell{r}{ 12.7 \\[-2pt]  (5.7) } \\
		&   & GRADE & \multirow{2}{*}{ \bigcell{r}{ 0.113 \\[-2pt]  (0.022) } } & \multirow{2}{*}{ \bigcell{r}{ 0.041 \\[-2pt]  (0.016) } } & \bigcell{r}{ 0.078 \\[-2pt]  (0.016) } & \bigcell{r}{ \textbf{0.002} \\[-2pt]  (0.001) } & \bigcell{r}{ \textbf{100} \\[-2pt]  (0.0) } & \bigcell{r}{ 34.5 \\[-2pt]  (5.1) } \\
		&   & SA-ODE &   &   & \bigcell{r}{ 0.071 \\[-2pt]  (0.047) } & \bigcell{r}{ 0.003 \\[-2pt]  (0.002) } & \bigcell{r}{ 98.7 \\[-2pt]  (3.9) } & \bigcell{r}{ \textbf{2.4} \\[-2pt]  (1.9) } \\
		\cmidrule{1-9}
		
		\multirow{3}{*}{ 100 } & \multirow{3}{*}{ 10 } & JADE & \bigcell{r}{ \textbf{0.328} \\[-2pt]  (0.065) } & \bigcell{r}{ \textbf{0.031} \\[-2pt]  (0.006) } & \bigcell{r}{ \textbf{0.147} \\[-2pt]  (0.073) } & \bigcell{r}{ 0.053 \\[-2pt]  (0.066) } & \bigcell{r}{ \textbf{96.9} \\[-2pt]  (7.9) } & \bigcell{r}{ 22.5 \\[-2pt]  (8.6) } \\
		&   & GRADE & \multirow{2}{*}{ \bigcell{r}{ 0.475 \\[-2pt]  (0.059) } } & \multirow{2}{*}{ \bigcell{r}{ 0.077 \\[-2pt]  (0.041) } } & \bigcell{r}{ 0.176 \\[-2pt]  (0.039) } & \bigcell{r}{ \textbf{0.002} \\[-2pt]  (0.002) } & \bigcell{r}{ 91.9 \\[-2pt]  (6.1) } & \bigcell{r}{ 23.2 \\[-2pt]  (6.0) } \\
		&   & SA-ODE &   &   & \bigcell{r}{ 0.226 \\[-2pt]  (0.152) } & \bigcell{r}{ 0.198 \\[-2pt]  (0.467) } & \bigcell{r}{ 98.1 \\[-2pt]  (6.7) } & \bigcell{r}{ \textbf{15.4} \\[-2pt]  (9.2) } \\
		\cmidrule{1-9}
		
		\multirow{3}{*}{ 100 } & \multirow{3}{*}{ 4 } & JADE & \bigcell{r}{ \textbf{1.794} \\[-2pt]  (0.449) } & \bigcell{r}{ \textbf{0.107} \\[-2pt]  (0.026) } & \bigcell{r}{ \textbf{0.401} \\[-2pt]  (0.271) } & \bigcell{r}{ 0.215 \\[-2pt]  (0.174) } & \bigcell{r}{ 95.7 \\[-2pt]  (8.4) } & \bigcell{r}{ 37.8 \\[-2pt]  (8.2) } \\
		&   & GRADE & \multirow{2}{*}{ \bigcell{r}{ 2.067 \\[-2pt]  (0.260) } } & \multirow{2}{*}{ \bigcell{r}{ 0.159 \\[-2pt]  (0.139) } } & \bigcell{r}{ 0.411 \\[-2pt]  (0.061) } & \bigcell{r}{ \textbf{0.004} \\[-2pt]  (0.004) } & \bigcell{r}{ 81.9 \\[-2pt]  (8.6) } & \bigcell{r}{ \textbf{13.5} \\[-2pt]  (3.2) } \\
		&   & SA-ODE &   &   & \bigcell{r}{ 0.630 \\[-2pt]  (0.313) } & \bigcell{r}{ 1.668 \\[-2pt]  (3.384) } & \bigcell{r}{ \textbf{97.6} \\[-2pt]  (5.8) } & \bigcell{r}{ 30.5 \\[-2pt]  (8.2) } \\
		\cmidrule{1-9}
		
		\multirow{3}{*}{ 40 } & \multirow{3}{*}{ 25 } & JADE & \bigcell{r}{ \textbf{0.125} \\[-2pt]  (0.025) } & \bigcell{r}{ \textbf{0.018} \\[-2pt]  (0.003) } & \bigcell{r}{ \textbf{0.116} \\[-2pt]  (0.055) } & \bigcell{r}{ \textbf{0.001} \\[-2pt]  (0.002) } & \bigcell{r}{ 89.9 \\[-2pt]  (9.4) } & \bigcell{r}{ 5.6 \\[-2pt]  (5.0) } \\
		&   & GRADE & \multirow{2}{*}{ \bigcell{r}{ 0.261 \\[-2pt]  (0.026) } } & \multirow{2}{*}{ \bigcell{r}{ 0.047 \\[-2pt]  (0.003) } } & \bigcell{r}{ 0.124 \\[-2pt]  (0.036) } & \bigcell{r}{ 0.003 \\[-2pt]  (0.002) } & \bigcell{r}{ \textbf{95.6} \\[-2pt]  (6.1) } & \bigcell{r}{ 32.2 \\[-2pt]  (5.6) } \\
		&   & SA-ODE &   &   & \bigcell{r}{ 0.120 \\[-2pt]  (0.064) } & \bigcell{r}{ \textbf{0.001} \\[-2pt]  (0.002) } & \bigcell{r}{ 89.4 \\[-2pt]  (10.9) } & \bigcell{r}{ \textbf{0.9} \\[-2pt]  (1.1) } \\
		\cmidrule{1-9}
		
		\multirow{3}{*}{ 40 } & \multirow{3}{*}{ 10 } & JADE & \bigcell{r}{ \textbf{0.614} \\[-2pt]  (0.100) } & \bigcell{r}{ \textbf{0.042} \\[-2pt]  (0.006) } & \bigcell{r}{ \textbf{0.190} \\[-2pt]  (0.089) } & \bigcell{r}{ 0.016 \\[-2pt]  (0.026) } & \bigcell{r}{ 85.8 \\[-2pt]  (14.8) } & \bigcell{r}{ 11.6 \\[-2pt]  (7.2) } \\
		&   & GRADE & \multirow{2}{*}{ \bigcell{r}{ 1.109 \\[-2pt]  (0.136) } } & \multirow{2}{*}{ \bigcell{r}{ 0.109 \\[-2pt]  (0.021) } } & \bigcell{r}{ 0.335 \\[-2pt]  (0.080) } & \bigcell{r}{ \textbf{0.010} \\[-2pt]  (0.011) } & \bigcell{r}{ \textbf{86.2} \\[-2pt]  (8.0) } & \bigcell{r}{ 19.7 \\[-2pt]  (5.6) } \\
		&   & SA-ODE &   &   & \bigcell{r}{ 0.305 \\[-2pt]  (0.221) } & \bigcell{r}{ 0.028 \\[-2pt]  (0.062) } & \bigcell{r}{ 83.9 \\[-2pt]  (13.9) } & \bigcell{r}{ \textbf{5.5} \\[-2pt]  (4.8) } \\
		\cmidrule{1-9} 
		
		\multirow{3}{*}{ 40 } & \multirow{3}{*}{ 4 } & JADE & \bigcell{r}{ \textbf{3.406} \\[-2pt]  (0.626) } & \bigcell{r}{ \textbf{0.146} \\[-2pt]  (0.026) } & \bigcell{r}{ \textbf{0.472} \\[-2pt]  (0.338) } & \bigcell{r}{ \textbf{0.072} \\[-2pt]  (0.092) } & \bigcell{r}{ 82.4 \\[-2pt]  (16.8) } & \bigcell{r}{ 20.0 \\[-2pt]  (12.3) } \\
		&   & GRADE & \multirow{2}{*}{ \bigcell{r}{ 4.735 \\[-2pt]  (0.966) } } & \multirow{2}{*}{ \bigcell{r}{ 0.285 \\[-2pt]  (0.098) } } & \bigcell{r}{ 0.762 \\[-2pt]  (0.472) } & \bigcell{r}{ 0.166 \\[-2pt]  (0.513) } & \bigcell{r}{ 75.6 \\[-2pt]  (9.5) } & \bigcell{r}{ \textbf{11.5} \\[-2pt]  (5.7) } \\
		&   & SA-ODE &   &   & \bigcell{r}{ 0.587 \\[-2pt]  (0.264) } & \bigcell{r}{ 0.238 \\[-2pt]  (0.390) } & \bigcell{r}{ \textbf{84.1} \\[-2pt]  (13.7) } & \bigcell{r}{ 13.3 \\[-2pt]  (9.1) } \\
		
		\bottomrule
	\end{tabularx}
	\label{tab:res_gaussian}
\end{table*}

Table~\ref{tab:res_gaussian} presents the results for Gaussian observations under various sample sizes and signal-to-noise ratios. In terms of latent process and derivative fitting, both GRADE and SA-ODE use the smoothing spline estimates without using any information from the ODE system. As expected, they are outperformed by JADE under all simulation settings since the estimates by JADE are regularized by the ODE fidelity and achieve substantial improvement. In some cases with a small sample size, JADE reduces about $30\%$ of $\textrm{MSE}(\wh\btheta)$ compared to the other competing methods.
In terms of estimating nonzero ODE additive components, JADE also demonstrates its strength. When the sample size is $40$ or $100$, JADE yields the smallest $\textrm{MSE}_{\textrm{active}}(\wh f)$, which is natural as JADE uses improved latent process estimates. However, when the smoothing spline estimates have already been accurate enough, the advantage of JADE over the two-stage methods becomes less significant. In particular, when the sample size is $200$, GRADE beats JADE in two experiments out of three, because it adopts an integrated form of the ODE system. Comparing the methods directly using the ODEs, JADE performs better than SA-ODE. Concerning the metric $\textrm{MSE}_{\textrm{inactive}}(\wh f)$, GRADE has a dominating performance because it yields fewer false positives, which is consistent with the findings in \citet{chen2017network}. For network recovery, all three methods can identify at least 80\% of the nonzero additive components correctly. Under the scenario with a small sample size and a lower noise level, the true-positive rates of the three methods can all reach 100\%. Roughly speaking, JADE and SA-ODE tend to select more nonzero additive components than GRADE.

\begin{table*}[htbp]
	\centering
	\caption{Poisson observation. Comparison of methods in latent process fitting, ODE additive components estimation and network discovery. Standard errors are displayed in the parentheses under the evaluation scores.}
	\ra{1.2}
	\tiny
	\begin{tabularx}{\textwidth}{@{}rrrRRRRRR@{}}
		\toprule
		Distribution & $n$ & Method 
		& \multicolumn{2}{c}{Latent Process} & \multicolumn{2}{c}{ODE Additive Component} & \multicolumn{2}{c}{Network Discovery} \\ 
		\midrule
		&  &  & \bigcell{c}{ $\textrm{MSE}(\wh\btheta)$ \\[-2pt] ($\times10^{-2}$)}   & $\textrm{MSE}(\wh \btheta' )$ & $\textrm{MSE}_{\textrm{active}}(\wh f)$ & $\textrm{MSE}_{\textrm{inactive}}(\wh f)$ & TP\% & FP\%  \\
		\cmidrule{4-9}
		
		\multirow{3}{*}{ Poisson } & \multirow{3}{*}{ 200 } & JADE & \bigcell{r}{ \textbf{0.193} \\[-2pt]  (0.037) } & \bigcell{r}{ \textbf{0.027} \\[-2pt]  (0.007) } & \bigcell{r}{ \textbf{0.134} \\[-2pt]  (0.047) } & \bigcell{r}{ 0.069 \\[-2pt]  (0.083) } & \bigcell{r}{ 99.4 \\[-2pt]  (2.7) } & \bigcell{r}{ 49.5 \\[-2pt]  (7.8) } \\ 
		&   & GRADE & \multirow{2}{*}{ \bigcell{r}{ 0.265 \\[-2pt]  (0.033) } } & \multirow{2}{*}{ \bigcell{r}{ 0.049 \\[-2pt]  (0.009) } } & \bigcell{r}{ 0.514 \\[-2pt]  (0.126) } & \bigcell{r}{ \textbf{0.055} \\[-2pt]  (0.017) } & \bigcell{r}{ \textbf{100} \\[-2pt]  (0.0) } & \bigcell{r}{ 42.1 \\[-2pt]  (4.1) } \\ 
		&   & SA-ODE &   &   & \bigcell{r}{ 0.346 \\[-2pt]  (0.241) } & \bigcell{r}{ 0.230 \\[-2pt]  (0.192) } & \bigcell{r}{ \textbf{100} \\[-2pt]  (0.0) } & \bigcell{r}{ \textbf{37.7} \\[-2pt]  (2.2) } \\ 
		\cmidrule{1-9} 
		
		\multirow{3}{*}{ Poisson } & \multirow{3}{*}{ 100 } & JADE & \bigcell{r}{ \textbf{0.364} \\[-2pt]  (0.066) } & \bigcell{r}{ \textbf{0.040} \\[-2pt]  (0.009) } & \bigcell{r}{ \textbf{0.159} \\[-2pt]  (0.070) } & \bigcell{r}{ 0.053 \\[-2pt]  (0.057) } & \bigcell{r}{ 99.5 \\[-2pt]  (2.6) } & \bigcell{r}{ 49.4 \\[-2pt]  (10.9) } \\ 
		&   & GRADE & \multirow{2}{*}{ \bigcell{r}{ 0.522 \\[-2pt]  (0.252) } } & \multirow{2}{*}{ \bigcell{r}{ 0.073 \\[-2pt]  (0.035) } } & \bigcell{r}{ 0.462 \\[-2pt]  (0.084) } & \bigcell{r}{ \textbf{0.041} \\[-2pt]  (0.014) } & \bigcell{r}{ \textbf{100} \\[-2pt]  (0.0) } & \bigcell{r}{ \textbf{36.7} \\[-2pt]  (3.8) } \\ 
		&   & SA-ODE &   &   & \bigcell{r}{ 0.425 \\[-2pt]  (0.434) } & \bigcell{r}{ 0.335 \\[-2pt]  (0.281) } & \bigcell{r}{ \textbf{100} \\[-2pt]  (0.0) } & \bigcell{r}{ 39.5 \\[-2pt]  (4.4) } \\ 
		\cmidrule{1-9} 
		
		\multirow{3}{*}{ Poisson } & \multirow{3}{*}{ 40 } & JADE & \bigcell{r}{ \textbf{0.988} \\[-2pt]  (0.404) } & \bigcell{r}{ \textbf{0.091} \\[-2pt]  (0.035) } & \bigcell{r}{ \textbf{0.233} \\[-2pt]  (0.069) } & \bigcell{r}{ 0.063 \\[-2pt]  (0.054) } & \bigcell{r}{ 96.8 \\[-2pt]  (6.6) } & \bigcell{r}{ 47.6 \\[-2pt]  (11.) } \\ 
		&   & GRADE & \multirow{2}{*}{ \bigcell{r}{ 1.306 \\[-2pt]  (0.636) } } & \multirow{2}{*}{ \bigcell{r}{ 0.127 \\[-2pt]  (0.037) } } & \bigcell{r}{ 0.445 \\[-2pt]  (0.144) } & \bigcell{r}{ \textbf{0.040} \\[-2pt]  (0.020) } & \bigcell{r}{ 99.5 \\[-2pt]  (2.5) } & \bigcell{r}{ 32.1 \\[-2pt]  (4.3) } \\ 
		&   & SA-ODE &   &   & \bigcell{r}{ 0.698 \\[-2pt]  (0.552) } & \bigcell{r}{ 1.867 \\[-2pt]  (2.269) } & \bigcell{r}{ \textbf{100} \\[-2pt]  (0.0) } & \bigcell{r}{ \textbf{39.8} \\[-2pt]  (4.9) } \\ 
		\bottomrule
	\end{tabularx}
	\label{tab:res_poisson}
\end{table*}

\begin{table*}[htbp]
	\centering
	\caption{Bernoulli observation. Comparison of methods in latent process fitting, ODE additive components estimation and network discovery. Standard errors are displayed in the parentheses under the evaluation scores.}
	\ra{1.2}
	\tiny
	\begin{tabularx}{\textwidth}{@{}rrrRRRRRR@{}}
		\toprule
		Distribution & $n$ & Method 
		& \multicolumn{2}{c}{Latent Process} & \multicolumn{2}{c}{ODE Additive Component} & \multicolumn{2}{c}{Network Discovery} \\ 
		\midrule
		&  &  & \bigcell{c}{ $\textrm{MSE}(\wh\btheta)$ \\[-0pt] ($\times10^{-2}$)}   & $\textrm{MSE}(\wh \btheta' )$ & $\textrm{MSE}_{\textrm{active}}(\wh f)$ & $\textrm{MSE}_{\textrm{inactive}}(\wh f)$ & TP\% & FP\%  \\
		\cmidrule{4-9}
		
		\multirow{3}{*}{ Bernoulli } & \multirow{3}{*}{ 200 } & JADE & \bigcell{r}{ \textbf{2.162} \\[-2pt]  (0.225) } & \bigcell{r}{ \textbf{0.176} \\[-2pt]  (0.011) } & \bigcell{r}{ \textbf{0.448} \\[-2pt]  (0.093) } & \bigcell{r}{ 0.108 \\[-2pt]  (0.084) } & \bigcell{r}{ \textbf{100} \\[-2pt]  (0.0) } & \bigcell{r}{ 46.9 \\[-2pt]  (6.9) } \\ 
		&   & GRADE & \multirow{2}{*}{ \bigcell{r}{ 2.185 \\[-2pt]  (0.225) } } & \multirow{2}{*}{ \bigcell{r}{ 0.179 \\[-2pt]  (0.009) } } & \bigcell{r}{ 0.587 \\[-2pt]  (0.140) } & \bigcell{r}{ \textbf{0.007} \\[-2pt]  (0.010) } & \bigcell{r}{ 91.5 \\[-2pt]  (8.3) } & \bigcell{r}{ \textbf{9.1} \\[-2pt]  (2.9) } \\ 
		&   & SA-ODE &   &   & \bigcell{r}{ 0.886 \\[-2pt]  (0.646) } & \bigcell{r}{ 0.323 \\[-2pt]  (0.238) } & \bigcell{r}{ 99.2 \\[-2pt]  (3.1) } & \bigcell{r}{ 35.8 \\[-2pt]  (2.0) } \\ 
		\cmidrule{1-9} 
		
		\multirow{3}{*}{ Bernoulli } & \multirow{3}{*}{ 100 } & JADE & \bigcell{r}{ \textbf{3.630} \\[-2pt]  (0.406) } & \bigcell{r}{ \textbf{0.217} \\[-2pt]  (0.016) } & \bigcell{r}{ \textbf{0.536} \\[-2pt]  (0.107) } & \bigcell{r}{ 0.115 \\[-2pt]  (0.057) } & \bigcell{r}{ \textbf{99.8} \\[-2pt]  (1.7) } & \bigcell{r}{ 46.7 \\[-2pt]  (5.8) } \\ 
		&   & GRADE & \multirow{2}{*}{ \bigcell{r}{ 3.645 \\[-2pt]  (0.405) } } & \multirow{2}{*}{ \bigcell{r}{ 0.221 \\[-2pt]  (0.017) } } & \bigcell{r}{ 0.628 \\[-2pt]  (0.100) } & \bigcell{r}{ \textbf{0.003} \\[-2pt]  (0.002) } & \bigcell{r}{ 89.4 \\[-2pt]  (7.6) } & \bigcell{r}{ \textbf{6.4} \\[-2pt]  (2.7) } \\ 
		&   & SA-ODE &   &   & \bigcell{r}{ 1.424 \\[-2pt]  (1.805) } & \bigcell{r}{ 0.503 \\[-2pt]  (0.523) } & \bigcell{r}{ 98.8 \\[-2pt]  (3.7) } & \bigcell{r}{ 35.9 \\[-2pt]  (2.3) } \\ 
		\cmidrule{1-9} 
		
		\multirow{3}{*}{ Bernoulli } & \multirow{3}{*}{ 40 } & JADE & \bigcell{r}{ \textbf{6.851} \\[-2pt]  (0.734) } & \bigcell{r}{ \textbf{0.285} \\[-2pt]  (0.024) } & \bigcell{r}{ \textbf{0.604} \\[-2pt]  (0.111) } & \bigcell{r}{ 0.146 \\[-2pt]  (0.116) } & \bigcell{r}{ \textbf{99.6} \\[-2pt]  (2.4) } & \bigcell{r}{ 47.3 \\[-2pt]  (7.7) } \\ 
		&   & GRADE & \multirow{2}{*}{ \bigcell{r}{ 6.967 \\[-2pt]  (0.675) } } & \multirow{2}{*}{ \bigcell{r}{ 0.292 \\[-2pt]  (0.025) } } & \bigcell{r}{ 0.791 \\[-2pt]  (0.134) } & \bigcell{r}{ \textbf{0.003} \\[-2pt]  (0.002) } & \bigcell{r}{ 79.0 \\[-2pt]  (8.4) } & \bigcell{r}{ \textbf{4.9} \\[-2pt]  (1.9) } \\ 
		&   & SA-ODE &   &   & \bigcell{r}{ 1.300 \\[-2pt]  (1.028) } & \bigcell{r}{ 0.975 \\[-2pt]  (0.666) } & \bigcell{r}{ 96.0 \\[-2pt]  (5.9) } & \bigcell{r}{ 35.3 \\[-2pt]  (2.3) } \\ 
		\bottomrule
	\end{tabularx}
	\label{tab:res_binomial}
\end{table*}

Table~\ref{tab:res_poisson} and \ref{tab:res_binomial} display the comparison results for Poisson or Bernoulli observations, respectively. Note that GRADE and SA-ODE are originally designed for Gaussian observations only. We extend them under the likelihood-based framework and develop an iteratively re-weighted least squares algorithm \citep{wood2017generalized} for efficient computation. Results of both Poisson and Bernoulli cases deliver a similar conclusion as the Gaussian case. Although we allow for multiple observations at each time point, it is in general a more challenging task for non-Gaussian data modeling. We notice that the estimation error and the false positive rates for network recovery are significantly higher than the Gaussian case. In terms of latent process and derivative fitting, JADE remains as the best because of the use of ODE regularization. Regarding the network recovery, GRADE delivers the best false positive rates even in the most challenging case with Bernoulli observations, while JADE and SA-ODE are conservative in selecting nonzero ODE components.

To sum up, GRADE adopts an integrated form of the ODE system and achieves the most reliable network recovery performance. JADE not only delivers competitive performance in network discovery, but is also superior in reducing the estimation error of latent process and additive ODE components.

\subsection{Insensitivity of $\lambda_\theta$}\label{sec:insenitive_lamtheta}

In Section~\ref{sec:tuning}, we claim that JADE is insensitive to $\lambda_\theta$ as long as it is not too small. We use the ODE system from Section~\ref{sec:setup} and generate $100$ Gaussian samples with the signal-to-noise ratio to be $10$ for each ODE process. All experiments are repeated for $100$ times, and the averaged results are given in Table~\ref{tab:lamtheta}. We observe that when $\lambda_\theta$ is no smaller than $0.1$, the performance of the JADE procedure has only minor variations. When $\lambda_\theta$ is smaller than $0.1$, the estimation of ODE additive components becomes worse, while other criteria remain robust. Therefore, we suggest that $\lambda_\theta$ can be fixed as a reasonable constant, for example, $\lambda_\theta=1$ in this small simulation.
\begin{table*}[h!]
	\centering
	\caption{Performance of JADE with different $\lambda_\theta$'s. Standard errors are displayed in the parentheses under the evaluation scores.}
	\footnotesize
	\begin{tabularx}{\textwidth}{cRRRRRR}
		\toprule
		$\lambda_\theta$ & \multicolumn{2}{c}{Latent Process} & \multicolumn{2}{c}{ODE Additive Component} & \multicolumn{2}{c}{Network Discovery} \\ 
		\midrule
		& \bigcell{c}{ $\textrm{MSE}(\wh\btheta)$ \\[-0pt] ($\times10^{-2}$)}   & $\textrm{MSE}(\wh \btheta' )$ & $\textrm{MSE}_{\textrm{active}}(\wh f)$ & $\textrm{MSE}_{\textrm{inactive}}(\wh f)$ & TP\% & FP\%  \\
		\cmidrule{2-7}
		0.01 & \bigcell{r}{ 0.323 \\[-1pt]  (0.043) } & \bigcell{r}{ 0.032 \\[-1pt]  (0.004) } & \bigcell{r}{ 0.194 \\[-1pt]  (0.081) } & \bigcell{r}{ 0.120 \\[-1pt]  (0.119) } & \bigcell{r}{ 100 \\[-1pt]  (0.0) } & \bigcell{r}{ 23.3 \\[-1pt]  (7.7) } \\ 
		0.1 & \bigcell{r}{ 0.355 \\[-1pt]  (0.059) } & \bigcell{r}{ 0.034 \\[-1pt]  (0.005) } & \bigcell{r}{ 0.154 \\[-1pt]  (0.065) } & \bigcell{r}{ 0.059 \\[-1pt]  (0.050) } & \bigcell{r}{ 100 \\[-1pt]  (0.0) } & \bigcell{r}{ 24.5 \\[-1pt]  (9.6) } \\ 
		1 & \bigcell{r}{ 0.347 \\[-1pt]  (0.055) } & \bigcell{r}{ 0.034 \\[-1pt]  (0.005) } & \bigcell{r}{ 0.161 \\[-1pt]  (0.063) } & \bigcell{r}{ 0.037 \\[-1pt]  (0.032) } & \bigcell{r}{ 100 \\[-1pt]  (0.0) } & \bigcell{r}{ 23.7 \\[-1pt]  (9.8) } \\ 
		10 & \bigcell{r}{ 0.333 \\[-1pt]  (0.051) } & \bigcell{r}{ 0.033 \\[-1pt]  (0.006) } & \bigcell{r}{ 0.162 \\[-1pt]  (0.067) } & \bigcell{r}{ 0.032 \\[-1pt]  (0.025) } & \bigcell{r}{ 100 \\[-1pt]  (0.0) } & \bigcell{r}{ 23.9 \\[-1pt]  (7.2) } \\ 
		100 & \bigcell{r}{ 0.324 \\[-1pt]  (0.050) } & \bigcell{r}{ 0.033 \\[-1pt]  (0.005) } & \bigcell{r}{ 0.149 \\[-1pt]  (0.058) } & \bigcell{r}{ 0.036 \\[-1pt]  (0.036) } & \bigcell{r}{ 100 \\[-1pt]  (0.0) } & \bigcell{r}{ 21.3 \\[-1pt]  (6.8) } \\ 
		\bottomrule	
	\end{tabularx}
	\label{tab:lamtheta}
\end{table*}

\section{Real Data Examples}\label{sec:data}

\subsection{Gene Regulatory Network in Yeast Cell Cycle}
Inferring gene regulatory networks (GRNs) has attracted many research interests \citep{emmert2014gene}. In this study, we focus on the GRN in the yeast cell cycle. It is beneficial to obtain a causal understanding of the regulatory effect of the genes controlled by the yeast cell cycle. The data used in our study is from the open database provided by \citet{spellman1998comprehensive}. They synchronize the cell cycle based on the $\alpha$-factor and measure the mRNA expression levels of 6,178 genes by the high-throughput microarray at 7-minute intervals for 119 minutes. We choose 100 representative genes that are identified to meet the minimum criterion for cell cycle regulation according to the study in \citet{spellman1998comprehensive}. For each gene, the time when the gene expression level reaches its peak is recorded. The peaking time may fall into one of the five phases of the cell cycle: G1, S, S/G2, G2/M, M/G1, indicating which period the gene displays its primary function. 

\begin{figure}[h]
	\centering
	\includegraphics[width=\linewidth]{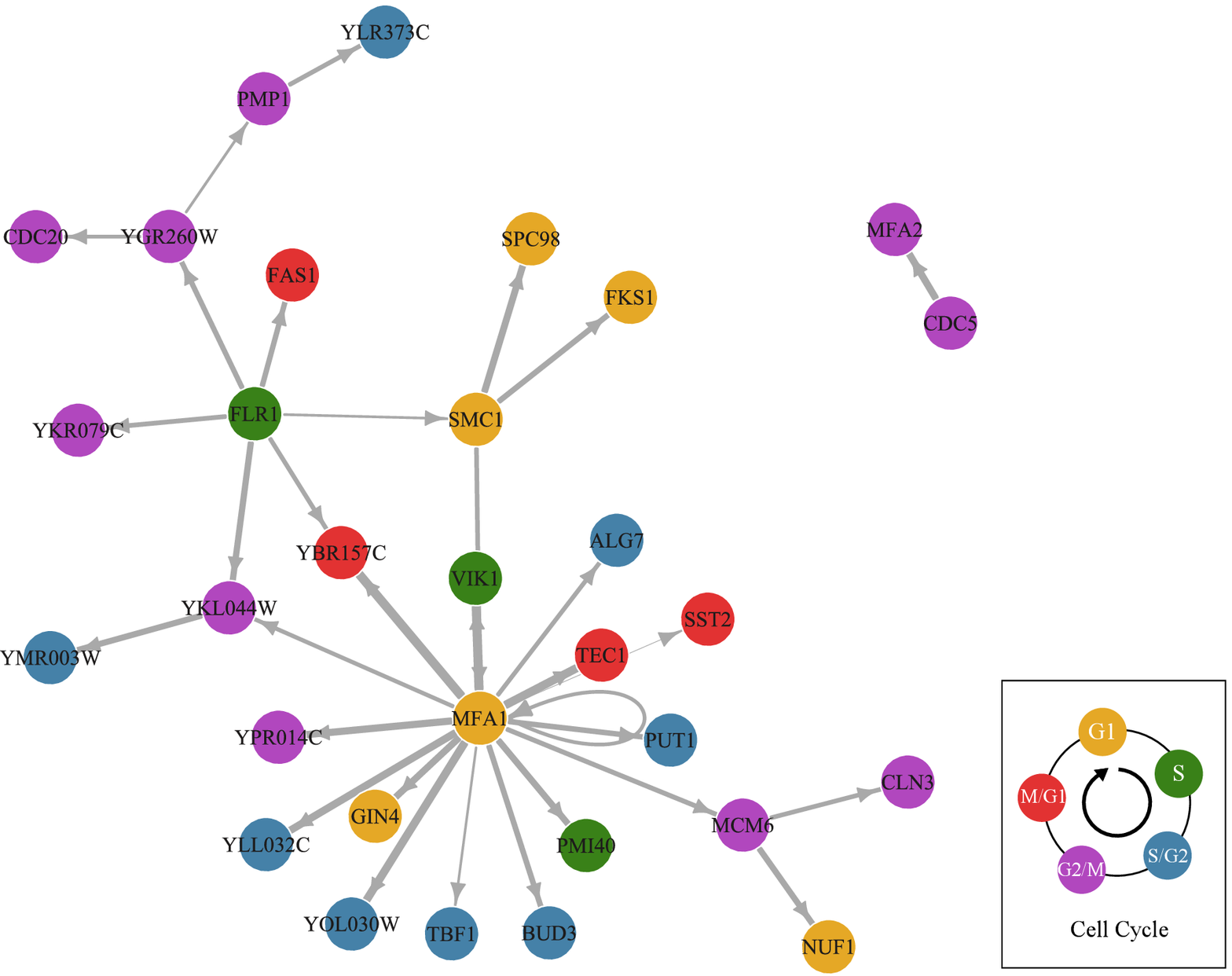}
	\caption{The estimated gene regulatory network among 100 genes during the yeast cell cycle, in which 63 isolated nodes are not displayed. Genes are shaded with different colors to indicate the phases where the expression levels reach their peaks.}
	\label{fig:grn}
\end{figure}

We treat the time-course gene expression levels as Gaussian observations and normalize them to zero mean and unit variance. The mean expression levels of the 100 genes are fitted as the latent ODE processes. The gene regulatory network recovered by JADE is displayed in Figure~\ref{fig:grn}. Among 100 genes, 37 genes are estimated to regulate or be regulated by other genes, while the rest 63 are regarded as isolated genes and are thus not displayed. Genes are shaded with different colors to indicate their corresponding cell cycle phase where the expression levels reach their peaks. The thickness of the edge indicates the strength of the estimated regulatory effect under $L_2$ norm.

Our result recognizes the MFA1 and FLR1 as the hub nodes that regulate the most genes. MFA1, an $\alpha$-factor precursor, has been identified to have a strong expression in the $\alpha$-factor based experiment \citep{spellman1998comprehensive}. It explains why it shows significant regulatory effects. FLR1 controls the production of a major facilitator superfamily that facilitates small solutes across cell membranes, which is also fundamental to biological processes \citep{nguyen2001multiple}.
We also find that some genes of the same peaking phase may form a connected cluster in the regulatory network, such as the cluster with SMC1, MFA1, FKS1 and SPC98, and the other cluster with YGR260W, CDC20 and PMP1. These clusters suggest there might be co-regulation among multiple genes during the same period.
We also notice that no gene of the S/G2 phase is connected with FLR1 that peaks at the S phase. Meanwhile, the S/G2 genes take about 35\% of all its regulated targets. This result implies that the formulation of the regulatory effect may depend on those phases.

We validate the reconstructed network with the Web-based Gene Set Analysis Toolkit (WebGestalt) \citep{liao2019webgestalt}. The WebGestalt performs enrichment analysis by first expanding the estimated network with 10 additional candidate genes besides the initial 100 genes and then generating a gene co-expression network, where the edges indicate strong correlations or dependencies in the expression. Details of the enrichment analysis is given in the Supplementary Material. By comparing the JADE and WebGestalt networks, we observe that the adjacent nodes identified by JADE are either directly connected or connected through one candidate gene at most in the WebGestalt network. For example, MFA1 and GIN4 both have co-expression with the candidate gene CCR4 in the WebGestalt network, which is represented by the edge between MFA1 and GIN4 in the JADE network.

\begin{figure}[h!]
	\centering
	\includegraphics[width=\linewidth]{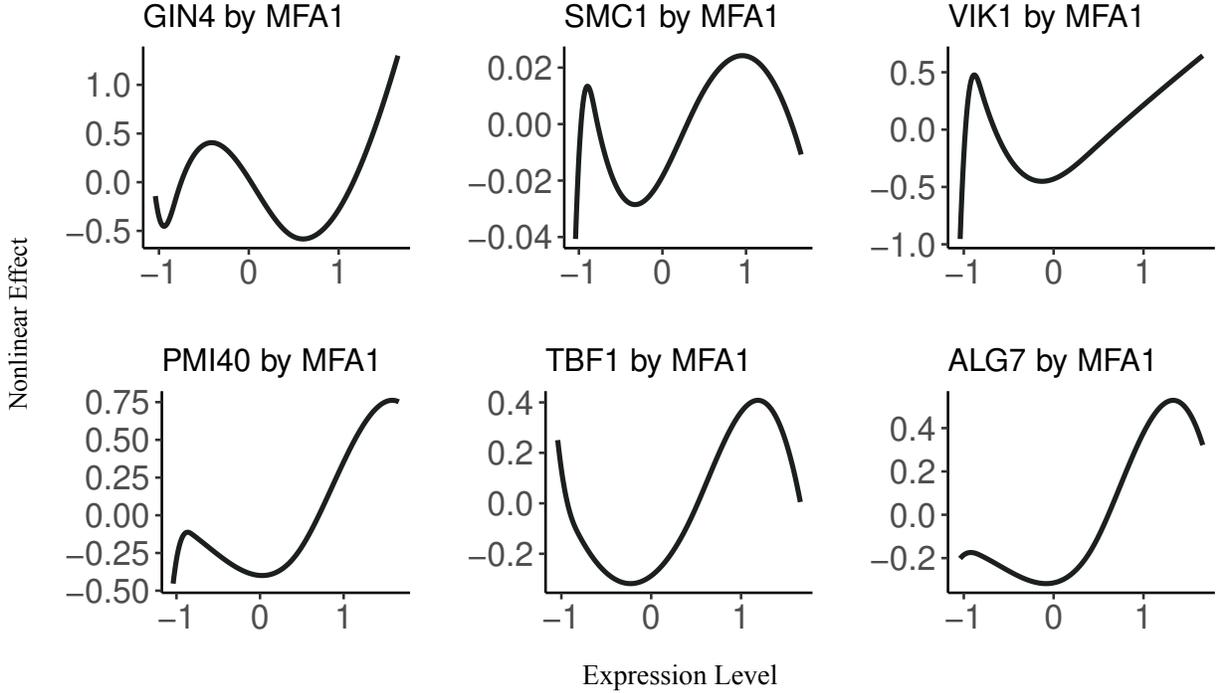}
	\caption{The nonlinear regulatory effects that the hub MFA1 exerts on six different genes. In each panel, the $x$-axis measures the mean expression level of MFA1, and the $y$-axis measures the instantaneous effect of MFA1 on the mean expression level of another gene.}
	\label{fig:grn_nonlinear_effect}
\end{figure}

Next, we inspect the nonlinear regulatory effects. In Figure~\ref{fig:grn_nonlinear_effect}, we display the regulatory effects of the hub MFA1 on six genes. All of them exhibit strong nonlinear patterns varying along with the gene expression level of MFA1. For example, the regulatory effect on PMI40 is negative when the expression level of MFA1 is low and is positive otherwise. This result implies the necessity of using the sparse additive ODE system for GRN modeling. It may help not only identify the nonlinear regulation effects, but also monitor the changing pattern of the effects along with the varying expression levels.

Comparison with the results by GRADE and SA-ODE is available in the Supplementary Material. All the three methods identify MFA1 and FLR1 as the hub nodes, confirming their vital regulatory functions. Moreover, the JADE network recognizes that FLR1 has a more influential role in this gene regulatory network. For example, FLR1 is connected to SMC1 which is related to an essential protein involved in chromosome segregation and double-strand DNA break repair \citep{cherry1998sgd, strunnikov1993smc1}. This finding suggests that the expression of SMC1 can be affected by the production of a major facilitator super-family controlled by FLR1. By contrast, neither the SA-ODE nor GRADE network has FLR1 connected to SMC1.

\subsection{Stock Price Change Direction}

The coronavirus pandemic, breaking out at the beginning of 2020, has resulted in severe economic disruption worldwide and implied massive turmoil on the financial market. To motivate the use of the JADE method in a practical application with Bernoulli observations, we consider modeling the change direction patterns of stock prices. The objective of this study is to estimate the latent probability processes of price increases and characterize the nonlinear influential effects among stocks. To this end, we select 40 companies and group them into $p=8$ sectors according to their major business. The number of companies in each sector is $L=5$. The list of the companies is presented in Table~\ref{tab:stocks}. 
\begin{table}[htbp]
	\footnotesize
	\centering
	\caption{Companies selected in eight sectors for stock price data analysis.}
	\begin{tabularx}{\linewidth}{ccX}
		\toprule
		\# & \textbf{Sector Name} & \textbf{Companies}  \\
		\midrule
		1 & \bigcell{c}{ Information \\[-1pt]Technology}  & Adobe, Apple, Microsoft, Salesforce, Zoom \\
		2 & \bigcell{c}{ Electric \\[-1pt]Vehicle} & BYD, Kandi, Nio, Tesla, Workhorse  \\
		3 & \bigcell{c}{Pharmaceutical} & AbbVie, Eli lilly, Moderna, Novartis, Pfizer \\
		4 & \bigcell{c}{Consumer\\[-1pt]Services \& Retail} & Ascena, J. C. Penney, Kohl's, Macy's, Nordstrom \\
		5 & \bigcell{c}{Online Retail\\[-1pt]Shopping} & Amazon, Best Buy, Target, Walmart, Wayfair \\
		6 & \bigcell{c}{Hotels} & Hilton, Marriott, Wyndham, Wynn, Park \\
		7 & \bigcell{c}{Air\\[-1pt]Transportation} & Boeing, Airbus, Delta Air Lines, Southwest Airlines, United Airline  \\
		8 & \bigcell{c}{Energy} & Chevron, Conocophillips, Exxon Mobil, Schlumberger, Valero Energy \\
		\bottomrule
	\end{tabularx}
	\label{tab:stocks}
\end{table}

\begin{figure}[h!]
	\centering
	\includegraphics[width=\linewidth]{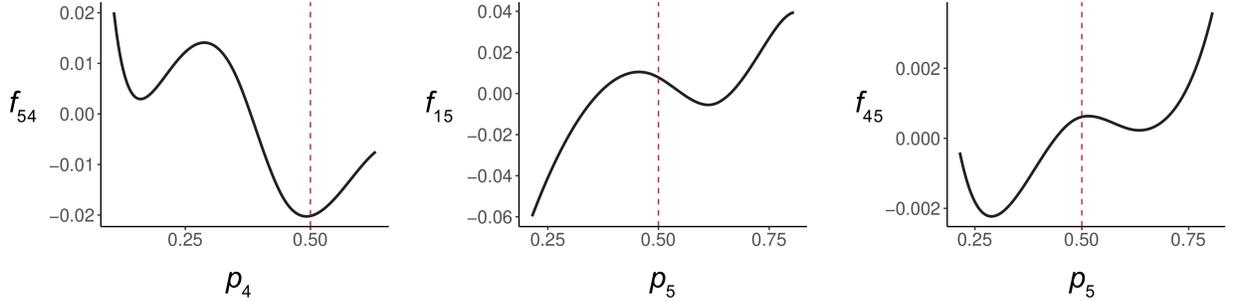}
	\caption{Three estimated nonlinear effects among eight sectors in the stock price data analysis. The $x$-axis $p_k, 1\leq k\leq 8,$ denotes the probability of stock price increase within Sector $k$, and the $y$-axis, denoted by $f_{jk}$, measures the effect of Sector $k$ on Sector $j$.}
	\label{fig:stock_nonlinear_effect}
\end{figure}

We collect the stock price indices for all trading days $t_1,\ldots, t_{248}$ during the year of 2020. For the $l$th company from Sector~$k$, where $1\leq l\leq 5$ and $1\leq k\leq 8$, we encode its stock price change direction on trading day $t_i$ as a Bernoulli observation. If its closing price on trading day $t_{i+d}$ is higher than that on trading day $t_i$, we set $y_{ikl}=1$ and $y_{ikl}=0$ otherwise. We choose $d=3$ to avoid unnecessary fluctuation in price changes. We assume that all stocks from the same sector share a common price changing pattern such that the Bernoulli observations are regarded as repeated measurements from a binomial distribution. More precisely, $y_{ik}=\sum_{l=1}^L y_{ikl}$ follows a binomial distribution with the probability of increase being $p_k(t_i) = 1/(1+\exp(-\theta_k(t_i)))$, where $\theta_k(t)$'s are the latent processes governed by a sparse additive ODE system
\begin{equation*}
	\theta^{'}_j(t) = \gamma_{j0} + \sum_{k=1}^8 f_{jk}(\theta_k(t)),\qquad j= 1,\ldots,8.
\end{equation*}

In Figure~\ref{fig:stock_nonlinear_effect}, we display three estimated nonlinear effects representing the potential interactions between sectors. In each panel, the $x$-axis measures the probability that stock prices in a sector will increase and the $y$-axis denoted by $f$ measures how the derivative of the other sector is affected. Given the $x$-coordinate, a positive $f$ value means there will be a higher probability of seeing the stock prices of the affected sector increase and a negative $f$ suggests the opposite tendency. In the top left panel, we observe that when the probability of an increasing price in Sector~4 (Consumer Services) is greater than $0.5$, the $f$ value is negative, which implies a slowing growth or a decreasing trend in the stock price of Sector~5 (Online Retail Shopping). Meanwhile, when the probability of price increase for Sector~4 is less than $0.1$, the $f$ value is positive, implying a potential price increase for Sector~5. Such a negative correlation indicates the competitive nature between Consumer Services and Online Retail Shopping. Due to the health concern brought by the coronavirus pandemic, the number of consumers visiting the department store significantly decreased. At the same time, innumerable online orders have boosted the stock price of e-commerce companies. Similarly, we observe upward momentum of Sector~5 (Online Retail Shopping) upon Sector~1 (Information Technology) and Sector~3 (Pharmaceutical). We also apply GRADE and SA-ODE to investigate the interactions among the same set of sectors and display them in the Supplementary Material. The estimated functional components are similar except in some regions. For example, $f_{54}$ estimated by GRADE wiggles around 0, indicating that Sector 5 (Online Retail Shopping) has no clear effect pattern on Sector 4 (Consumer Services); $f_{15}$ estimated by SA-ODE has a sharp change from positive to negative in $[0.4,0.6]$ and then quickly increases from negative to positive in $[0.6,0.8]$, which seems unrealistic.
\begin{figure}[h!]
	\centering
	\includegraphics[width=\linewidth]{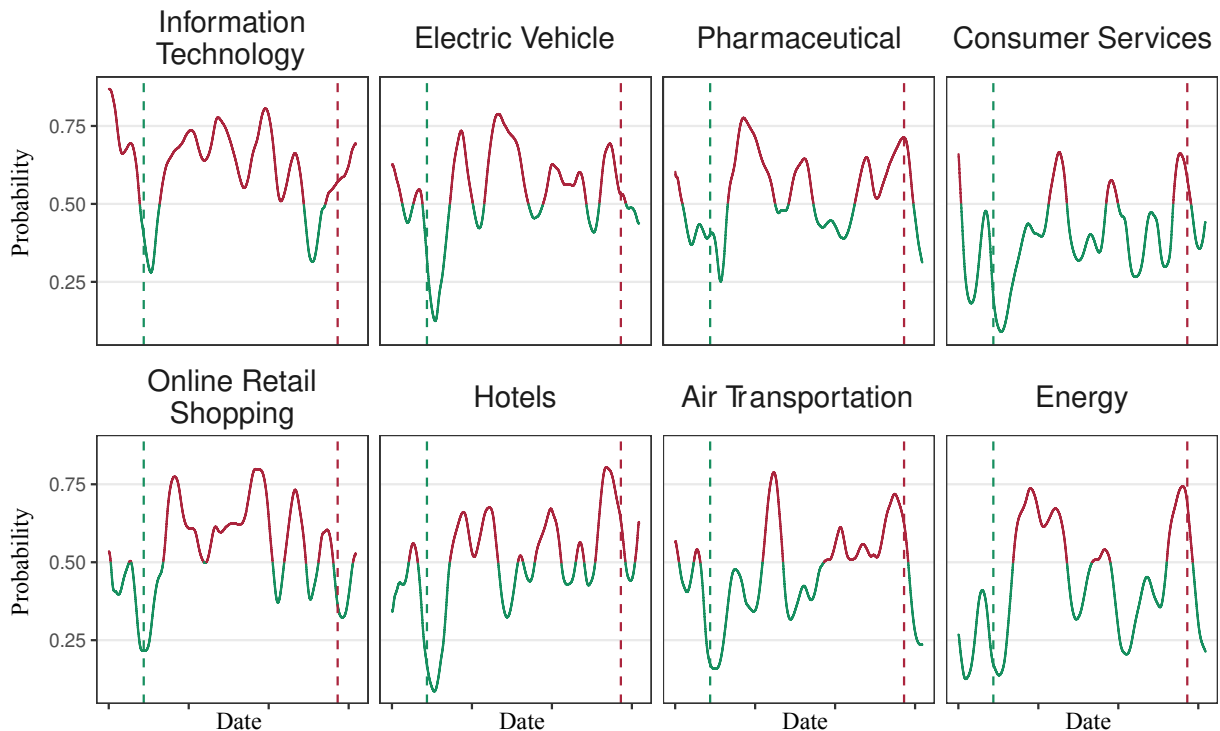}
	\caption{The fitted probabilities that the stock prices will increase in three trading days varying along the time. Each trajectory starts from January 1 to December 31 in 2020. The red color indicates the probability is above $0.5$, and the green color indicates the opposite. The green dashed lines mark February 20 and the red dashed lines mark November 24.}
	\label{fig:stock_trajectories}
\end{figure}

The estimated ODE trajectories for each sector are plotted in Figure~\ref{fig:stock_trajectories}. We first observe that all the trajectories dropped to their valleys after the sudden crash of the global stock market on February 20, 2020. During that period, multiple circuit breakers were triggered by the coronavirus pandemic. From then on, several sectors kept suffering from the long-lasting stock price decrease, such as the consumer services and the energy industry. In the meantime, some other companies grabbed the opportunities and made huge profits. Besides the online retail companies, information technology companies kept boosting under the growing demands for information services and electronics devices. At the end of the year 2020, we witnessed rebounds in all sectors. On November 24, Dow Jones hit 30,000 during the United States presidential transition.

\section{Conclusion}\label{sec:conclude}
In this article, we propose a new approach called JADE for parameter estimation in generalized sparse additive ODEs. A unified estimation framework is developed by taking into account the data likelihood, ODE fidelity and sparse regularization simultaneously. It covers existing two-stage collocation methods, the generalized profiling method and our JADE procedure. As a joint approach, JADE has the advantage in estimating the latent processes and additive ODE components by regularizing the smooth fits with the ODE system, while the group lasso penalty helps achieve a sparse estimation of the network structure at the same time. 

To address the computational issue, we design a block coordinate descent algorithm with provable global convergence. Compared to the classical iteratively re-weighted least squares algorithm for generalized linear models, JADE adopts a gradient-descent optimization strategy where an inexact line search is applied instead of exact Hessian calculations. It is simple, highly parallelizable, and achieves great generalization performance. Empirically, JADE is superior in both latent process and ODE parameter estimation together with reliable sparse network identification. To sum up, JADE is a good alternative for dynamical modeling of non-Gaussian data with nonlinear ODEs.


\appendix
	
\section{Line Search with the Armijo Rule}\label{sec:Armijo}
	
	In Section~\ref{sec:bcd}, we discussed gradient descent methods with step sizes selected by the Armijo rule. Following \citet{tseng2009coordinate}, we present a necessary background on this subject. 
	
	Let $Q(\bbeta)=S(\bbeta)+R(\bbeta)$ be the objective function to be minimized, where $S$ is smooth on an open subset of $\mathbb R^p$, and $R$ is convex and possibly non-smooth. Given $\alpha_{\rm init}>0$, the Armijo rule selects the step size $\alpha$ to be the largest element among the sequence $\{\alpha_{\rm init} \eta^j: j=0,1,\dots\}$ satisfying
	\begin{equation}\label{eq:armijo}
		\begin{aligned}
			Q(\bbeta+\alpha \bd)&\leq Q(\bbeta)+\alpha\zeta\bigg[\{\nabla S(\bbeta)\}^\top \bd
			+\kappa \bd^\top \bH \bd + R(\bbeta+\bd)-R(\bbeta)\bigg],
		\end{aligned}
	\end{equation}
	where $0<\eta<1$, $0<\zeta<1$, $0\leq \kappa<1$, and $\bH\in\mathbb R^{p\times p}$ is a positive definite matrix.

	\section{ODE Specification in Simulation Study}\label{sec:ode_appendix}
	
\begin{figure}[ht]
		\centering
		\includegraphics[width=\linewidth]{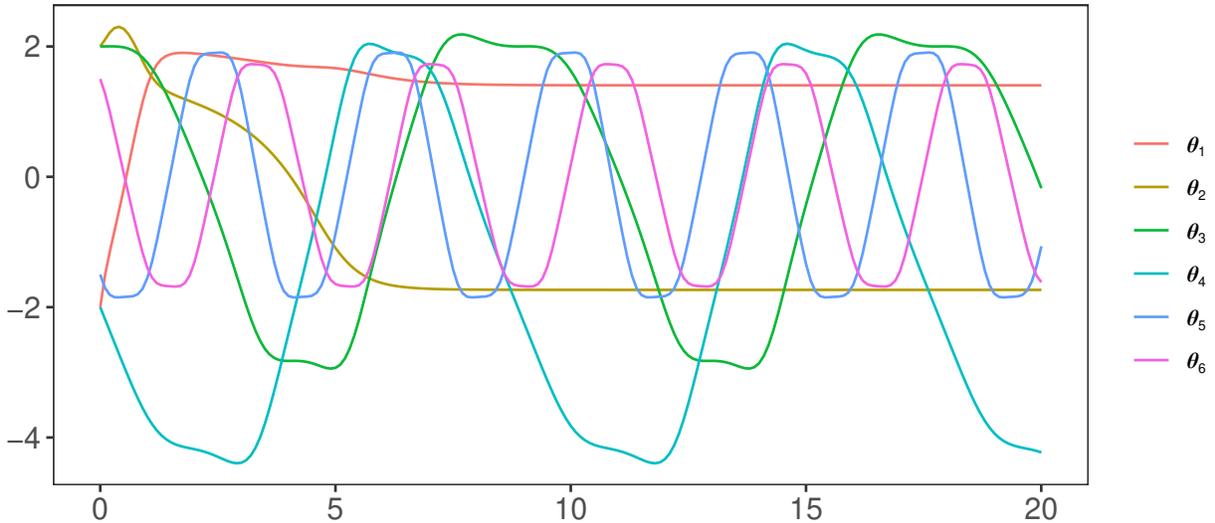}
		\caption{The first six generated ODE solutions used in the numerical study.}
		\label{fig:ode}
	\end{figure}

	In Section~\ref{sec:setup}, we use \eqref{eq:ode_add_gen} to generate pairs of ODE solutions on the interval $\mathcal{T}=[0,20]$. The detailed ODE parameter specification is as follows:
	\begin{enumerate}
		\item $\theta_1(t)$, $\theta_2(t)$ are generated with $\beta_{1,0} = 0$,  $\bbeta_{1,1} = (1.2, 0.3, -0.6)^\top$,
		$\bbeta_{1,2} = (0.1, 0.2, 0.2)^\top$, $\beta_{2,0} = 0.4$, $\bbeta_{2,1} = (-2, 0, 0.4)^\top$, $\bbeta_{2,2} = (0.5, 0.2, -0.3)^\top$.
		
		\item $\theta_3(t)$, $\theta_4(t)$ are generated with $\beta_{3,0} = -0.2$,  $\bbeta_{3,3} = (0, 0, 0)^\top$,
		$\bbeta_{3,4} = (-0.3, 0.4, 0.1)^\top$, $\beta_{4,0} = -0.2$, $\bbeta_{4,3} = (0.2, -0.1, -0.2)^\top$, $\bbeta_{4,4} = (0, 0, 0)^\top$.
		
		\item $\theta_5(t)$, $\theta_6(t)$ are generated with $\beta_{5,0} = 0.05$,  $\bbeta_{5,5} = (0, 0, 0)^\top$,
		$\bbeta_{5,6} = (0.1, 0, -0.8)^\top$, $\beta_{6,0} = -0.05$, $\bbeta_{6,5} = (0, 0, 0.5)^\top$, $\bbeta_{6,6} = (0, 0, 0)^\top$.
		
		\item $\theta_7(t)$, $\theta_8(t)$, $\theta_9(t)$, $\theta_{10}(t)$ are generated with constant derivatives, where $\beta_{7,0}$, $\beta_{8,0}$, $\beta_{9,0}$, $\beta_{10,0}$ are sampled from the standard normal distribution.
	\end{enumerate}
	Figure~\ref{fig:ode} displays the first six ODE solutions to the above ODE system, while the rest four are linear in $t$.
	When generating Gaussian samples, we can directly use these trajectories as the latent processes. When generating Poisson and Bernoulli samples, it may cause some troubles with a direct application. More precisely, it is hard to accurately estimate $\theta(t)$ when the intensity $\exp(\theta(t))$ is very small. Therefore, we rescale $\theta_j(t)$ to $(\theta_j(t) - b_j)/a_j$ such that the intensity $\exp\{\theta_j(t)\}$ for Poisson distribution or the probability $1/[1 + \exp\{-\theta_j(t)\}]$ for Bernoulli distribution is well spread over its range. The specific values for $a_j$ and $b_j$ are given in Table~\ref{tab:ajbj}.
	\begin{table}[htbp]
		\centering
		\caption{Constants $a_j$ and $b_j$ used in the linear transformation of $\theta_j(t)$ when generating Poisson or Bernoulli samples, where $m_j=\min_{t\in\mathcal{T}}\{\theta_j(t)\}$ and $M_j=\max_{t\in\mathcal{T}}\{\theta_j(t)\}$. }
		
		\begin{tabular}{@{}c cc cc@{}}
			\toprule
			\multirow{2}{*}{$j$} & \multicolumn{2}{c}{Poisson} & \multicolumn{2}{c}{Bernoulli} \\
			\cmidrule(lr){2-3} \cmidrule(lr){4-5}
			& $a_j$ & $b_j$ & $a_j$ & $b_j$ \\
			\midrule
			$1,2$ & $1$ & $m_j-1$ & $0.2\,(M_j-m_j)$ & $0.5\,(m_j + M_j)$ \\
			$3,4$ & $1.5$ & $m_j-1$ & $0.2\,(M_j-m_j)$ & $0.5\,(m_j + M_j)$ \\
			$5,6$ & $1$ & $m_j-1$ & $0.2\,(M_j-m_j)$ & $0.5\,(m_j + M_j)$ \\
			\bigcell{c}{$7,8$, \\[-1pt] $9,10$} & $1$ & $m_j-0.1$ & $0.2\,(M_j-m_j)$ & $0.5\,(m_j + M_j)$ \\
			\bottomrule
		\end{tabular}
		\label{tab:ajbj}
	\end{table}

\bigskip
\begin{center}
{\large\bf SUPPLEMENTARY MATERIALS}
\end{center}

\begin{description}

\item[Supplementary Material] contains additional numerical experiments.

\end{description}

\bibliographystyle{Chicago}

\bibliography{bib-ode-exp-np}
\end{document}